\documentclass[12pt]{article}
\usepackage[dvips]{graphicx}
\usepackage{amsmath,amssymb}

\begin{document}
\title{ On coherence lengths of  wave packets II : High energy neutrino } 
\author{ K. Ishikawa and  Y. Tobita}
\maketitle
\begin{center}
  Department of Physics, Faculty of Science, \\
Hokkaido University Sapporo 060-0810, Japan
\end{center}
\begin{abstract}
In this second paper of the series on the coherence of wave packets, 
we study neutrinos in high energy experiments where
 neutrinos are produced by decays of  pions or muons which are described
 using  wave packets. 
The space time
 position where one neutrino is   produced is not fixed to one value but
 is extended in macroscopic area.  Hence the amplitude is 
 defined by a superposition of the amplitudes of different neutrino's
production time in the macroscopic region and depends on the absolute value
 of the neutrino mass.

We analyze  neutrino interference based on operator product expansion
 near the light cone and  find a  new universal  term 
in the time dependent neutrino probability. This new term  has an origin in
 higher order quantum effect in a similar manner as axial anomaly.  Roles of  
Lorentz invariance and the operator product expansion in the light-cone
 region are clarified and a  possibility of measuring the absolute value  
of neutrino mass from neutrino interference experiments is pointed out.

\end{abstract}
\newpage

\section{Introduction}
One particle states in nature are described by wave packets in various
situations. In this series of papers
\cite{Ishikawa-Shimomura,Ishikawa-Tobita-ptp,Ishikawa-Tobita}, we study
physics and implications of wave packets. By using wave packets,
space-time dependent amplitudes and probabilities  that are  impossible to
obtain using standard scattering amplitudes are computed. The time
dependent probability thus obtained is shown to  give a valuable
information on the 
absolute value of neutrino mass.

Neutrinos are particles that are very light and interact with
matters very weakly. Their  masses were found to be finite from recent 
flavour oscillation
experiments~\cite{SK-Atom,SK-Solar,SNO-NC,KamLAND-Reactor,Borexino,K2K}. 
The neutrino oscillation experiments using
neutrinos from the sun,
accelerator, reactors, and atmosphere  gave the values of differences of 
the mass squared. Their values are found to be \cite{particle-data}
\begin{eqnarray}
& &\Delta m^2_{21} = m_2^2 - m_1^2 = (0.8 \pm 0.03)\times 10^{-4}~[\text{eV}^2/c^4]\\
& &|\Delta m^2_{32}| = |m_3^2 - m_2^2|= 0.19 \ \text{to} \ 0.3\times 10^{-2}~[\text{eV}^2 /c^4],
\end{eqnarray}
with certain uncertainties where $m_i~(i=1-3)$ are mass values.
The  squared-mass  differences  are  extremely small but the absolute
values of masses are unknown. Tritium beta decays \cite{Tritium}~have
been used for determining the 
absolute value but the existing upper bound for the effective electron 
neutrino squared-mass  is of order  $  2 ~(\text{eV}/c^2)^2$ and the mass 
is $ 0.2-2 ~\text{eV}/c^2$ from cosmological observations~\cite{WMAP-neutrino}. 
Neutrino masses are far from other particle's masses and are important
parameters of physics. Neutrino masses are not affected from standard
model  of electroweak gauge interactions and it is important to know 
precise values of neutrino masses.    

Neutrinos interact with matter by  weak interactions and 
event rates are very low and neutrino detection is hard. 
However using its weak interactions with matters, neutrinos can be used
as new observational means  once the detection method is
established~\cite{KamLAND-Geo}. Using  neutrinos, several astronomical objects
such as sun, moon, and other stars inside of which can not be observed
directly by ordinary means such as lights, electrons and protons, would 
be  studied in a future~\cite{Ishikawa-shimomura-lunar}. For these applications, it is necessary 
 to know  precise  properties of neutrinos.  We study wave and 
particle properties of  high energy neutrinos. 
  
A double slit-like interference of a neutrino, which is totally different from
the flavour oscillation, is a subject of the present work.        
Neutrinos are produced by weak decays of  particles and propagate finite
distance before it is detected.  The distance is not fixed but varies within 
  certain range. So the wave at the detector is a superposition of the
  neutrino produced at different positions.  We study the
pion and muon decays and the neutrino
detection probability in hadron collisions using wave packets. 
The scattering amplitude
and probability are computed usually using plane waves but the finite
time or finite distance behaviors are not known by the standard method 
and wave packets are suitable means to study these dependences.

When the 
particles involved in a reaction are  wave packets of  finite  spatial 
extensions and the particle measurement is made by wave packet, the time 
and space dependent transition probability can be  observed. By using
wave packets, the amplitude and
probability of the finite time
interval are  also calculated.   
The probability  obtained by  our calculation has  an oscillating term 
in time interval, T, in addition to the normal T-linear term. The normal term  
is calculable also in the standard S-matrix in the momentum
representation, but  the anomalous
T-oscillating term in the finite time is calculable only using wave packet.
The wave packet formalism supplies the
space-time dependent informations and we obtain  the length   dependent
probability of 
finding neutrino. This dependence is generated since
the position where the neutrinos are produced in the decay of hadrons is 
not fixed and extended to macroscopic area.
The neutrino amplitude is a superposition of the amplitude of 
different time and position and shows the interference.

We combine pion dynamics with the weak interaction of the
neutrino and find the time
dependent  amplitude and probability.  
This amplitude shows  neutrino interferences  of
anomalous behavior that is sensitive to the neutrino mass. 
The probability oscillates with  time in a scale that
is much shorter than that of flavour oscillation. 
The wave length  of this oscillation   is determined by the
energy and mass of neutrino and  the interference experiments could be
useful for  finding  the absolute value of the neutrino mass.  

We have shown  general 
features of wave packet scattering in  \cite{Ishikawa-Shimomura} and 
of particle coherence in the previous paper I
\cite{Ishikawa-Tobita-ptp} \footnote{The general arguments about
the wave packet scattering are in \cite{Goldberger,newton,Sasakawa}.}. An 
 important 
feature of the wave packet of the relativistic particle is that the
phase factor of the
wave function is determined by the mass and the energy in a relativistic
invariant manner.
For the neutrino
of mass $m_{\nu}$ and energy $E_{\nu}$ the phase factor is expressed
by using the differences of two positions $\Delta {\vec x}={\vec x}-{\vec X}$
and of two times
$\Delta t=t-T$, as $\exp{(i \phi)}$, where the phase $\phi$ is expressed
as $\phi=m_{\nu}\sqrt{c^2{\Delta t}^2-{\Delta x}^2} $ and also is written as 
$\phi={m_{\nu}^2 \over  E_{\nu}} \times  c \Delta t$, where $(t,{\vec x})$ are time and space
coordinates of the production point and $(T,{\vec X})$ are those of the
detection point. Consequently the interference due to this phase   
is also determined by the energy and mass. For the neutrino of very small mass,
of order $1$~eV or
less,  this interference is in the large scale. 

We
investigate the  physical problems  that are connected with neutrino's 
wave packets and interferences in high energy regions. 
Particularly neutrinos from pion decay and muon decay are
studied in this paper.
Other low energy neutrino processes caused by solar neutrinos, reactor
neutrinos, and others are studied in a next
work.

This paper is organized in the following manner. In section 2, wave
packet sizes of decaying particles are estimated.  In section  3, we
study  neutrino  production amplitude  and probability in hadron collisions and 
in section 4 we study neutrinos from real pions decays and those of 
 muon decay in section 5.  
Summary and prospects are given in section 6.

\section{Wave packet sizes}
When decaying particles are not exact plane waves but are the  wave
packets of finite coherence lengths, the produced particles have also 
these properties of coherence. We estimate coherence lengths of
proton first and those of pion and muon next following the methods of
our previous works  \cite{Ishikawa-Shimomura} 
\cite{Ishikawa-Tobita-ptp}.

\subsection{Pion  wave packets }
Pions are produced by the proton collisions with target nucleus. Hence
the coherence property of pion is determined by the coherence property  of
proton and nucleus.  The proton has a finite scattering probability with nucleus in
the matter and has a finite coherence length and the target nucleus
has a microscopic size of order $10^{-15}$~m and its position is fixed with the
uncertainty of atomic distance in crystals. From these values 
coherence length of proton is determined and using the proton coherence length
we estimate the pion coherence length. 

\subsubsection{Proton  mean free path}

As was shown  in I \cite{Ishikawa-Tobita-ptp}, the coherence length of a
particle in dense matter is 
determined by its mean free path. High energy 
particles in solid  material interact with atoms frequently and 
its  mean free path is understood well.  
The mean free path is an average distance for one particle to move freely 
and maintains  particle's  coherence. A particle is expressed by one wave
function in a finite distance defined by the mean free path. Beyond the mean 
free path, particles lose coherence and are expressed by a different wave
function. Hence  this particle state is different from the plane
wave. The wave function which expresses this particle has a finite
spatial size and  finite  momentum width. 

The mean free path of the charged particle 
is determined by its  scattering with atoms in matter by Coulomb
interaction. The energy loss is also determined by the same cross
section  and  is summarized well  in particle data summary
\cite{particle-data}. 
In order to estimate a mean free path of the proton of $1~\text{GeV}/c$, we use 
the proton energy loss rate for several metals such as Pb, Fe, and others as,
\begin{eqnarray}
-{d E \over d x}=1-2 ~[\text{MeVg}^{-1}\text{cm}^2],   
\end{eqnarray}
we have the mean free path of the $1~\text{GeV}/c$ proton  
\begin{eqnarray}
L_\text{proton}={E \over {dE \over d x}\times \rho }= {1 ~[\text{GeV}] \over (1-2)\times
 10 ~[\text{MeV g}^{-1} \text{cm}^2 \text{g cm}^{-3}]} = 50-100~[\text{cm}].
\end{eqnarray}
At lower energy of $2~\text{MeV}/c$, the energy loss rate is about
$10~\text{MeVg}^{-1}\text{cm}^2$ and the mean free path is  
\begin{eqnarray}
L_\text{proton}=10~[\text{cm}].
\end{eqnarray}

Actually proton which is accelerated and reaches high energy at the end
has a size of wave packet after  acceleration, $L_{after}$,
\begin{eqnarray}
L_\text{after}= L_\text{before}\times{v_\text{after} \over
 v_\text{before}},
\end{eqnarray}
where  $L_{before }$ is the size before the acceleration and $v_{before }$
and $v_{after}$ are the
velocities  before and after the acceleration. The velocity is bounded
by the light velocity $c$,
and the ratio for reaching from the above momenta to $10~\text{GeV}/c$
 is from about $1.2$ at $1~\text{GeV}/c$ to five at $0.2~\text{GeV}/c$. Hence the proton
 of $10~\text{GeV}/c$ has the mean free path
\begin{eqnarray}
L_\text{proton}\approx 40 - 100~[\text{cm}].
\label{meanfp-.2Gev}
\end{eqnarray}

\subsubsection{Pion mean free path }

Pions are produced by a collision of the proton with target nucleus. Coherence 
lengths of pions which are produced  in high energy proton 
collisions with target nucleus are obtained    using the above initial 
proton coherence  size and  target size. In relativistic energy region,
particles have light velocity.   Hence in the pion production, the
coherence length of the pion ${\delta x}_f$ is given from that of the proton
${\delta x}_i$ as,
\begin{eqnarray}
& &{{\delta x}_i \over v_i}={{\delta x}_f \over v_f },\\
& &{\delta x}_f={v_f \over v_i}{\delta x}_i\approx {\delta x}_i .
\nonumber
\end{eqnarray}
Consequently from Eq.$(\ref{meanfp-.2Gev})$,  the pion's coherence
length of 
the momentum $1~\text{GeV}/c$ or larger 
momentum is given by
\begin{eqnarray}
L_\text{pion}\approx 40 -100~[\text{cm}].
\label{mean-free-pion}
\end{eqnarray}   
We use these values of Eq.~$(\ref{meanfp-.2Gev})$ and 
Eq.~$(\ref{mean-free-pion})$ in latter sections.  

\subsection{Muon   wave packet}
The muon is produced from pion decay. By the decay of pion of finite 
coherence length, a finite coherence length of produced muon is
determined. 
\subsubsection{Decay of pion}
Coherence lengths of the muon is connected with that of the pion by the
ratio of velocities,
 \begin{eqnarray}
{{\delta x}_\text{pion} \over v_\text{pion}}={{\delta x}_\text{muon} \over v_\text{muon}},
\end{eqnarray}
and is expressed as
\begin{eqnarray}
\delta x_\text{muon}={v_\text{muon} \over v_\text{pion}}\times \delta x_\text{pion}.
\label{meanfp-muon-ratio}
\end{eqnarray}
For the relativistic particles the velocities are light velocity and 
the velocity ratio is unity.
 
Since the initial pion has an extension of momentum ${\Delta p}_\text{pion}$
the final
muon has also an extension of momentum ${\Delta p}_\text{muon }$,  
\begin{eqnarray}
{\Delta p}_\text{muon}={\Delta p}_\text{pion}+O({\hbar \over \delta x}_i).
\end{eqnarray}

\subsubsection{Muon coherence length}
Combining Eq.~(\ref{mean-free-pion}) and Eq.~(\ref{meanfp-muon-ratio}),
the coherence length of muon is given by 
\begin{eqnarray}
L_\text{muon}\approx 40 - 100~[\text{cm}].
\label{mean-free-muon}
\end{eqnarray}  
\subsection{Neutrino    wave packet}
The size of wave packet for observed neutrino  is determined by the
object that neutrino
interacts  in  detectors. Neutrinos interact with nucleus or with
electrons in atoms.  The nucleus have sizes of order $10^{-15}$~m and the
electron's wave functions have sizes of order $10^{-10}$~m. 
We study the neutrinos
described by the wave packets of these sizes in many particle
processes. In this respect, the
neutrino wave packet of the present work is different from some  previous
works of wave packets that are connected with flavour neutrino
oscillations~\cite{Kayser,Giunti,Nussinov,
Kiers,Stodolsky,Lipkin,Asahara}, where one particle properties of
neutrino at production are studied. It is important to study the
neutrino wave packet at the detector to study the interference.

\begin{figure}[t]
 \includegraphics[scale=1.2]{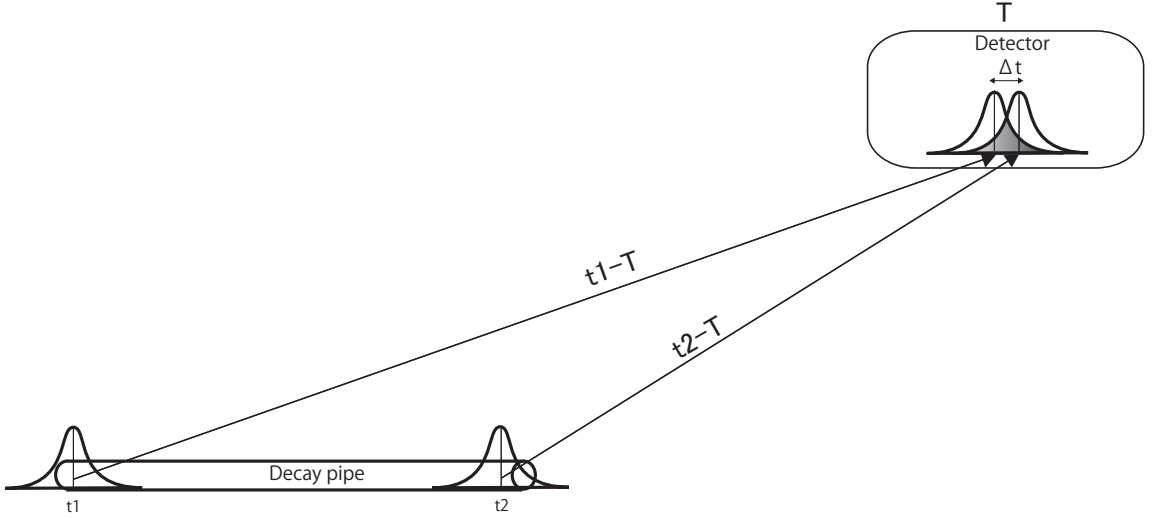}
\caption{The geometry of the neutrino interference experiment. The
 neutrino is observed by the detector at T and produced at $t_1$ or $t_2$.}
\label{fig:geo}
\end{figure}

The muon neutrino interactions in detectors are 
\begin{eqnarray}
& &\nu_{\mu}+e^{-} \rightarrow e^{-}+\nu_{\mu} 
\label{numu-leptonic}\\
& &\nu_{\mu}+e^{-} \rightarrow \mu^{-}+\nu_{e} 
\label{numu-lcharged}\\ 
& &\nu_{\mu}+A \rightarrow \mu^{-}+(A+1)+X 
\label{numu-hcharged}\\ 
& &\nu_{\mu}+A \rightarrow \nu_{\mu} +A  +X
\label{numu-hneutral}
\end{eqnarray}
The neutrino 
wave packet  $\sigma_{\nu}$ in processes $(\ref{numu-leptonic})$ and
$(\ref{numu-lcharged})$ is of order   $10^{-10}$~m and the neutrino 
wave packet  $\sigma_{\nu}$ in processes $(\ref{numu-hcharged})$ and
$(\ref{numu-hneutral})$ is of order   $10^{-15}$~m. In the  following
sections are discussed  the neutrinos in short or intermediate baseline 
experiments. We
will see that the neutrino production amplitudes are unchanged even
though the smaller wave packet, $10^{-15}$, is used.  The reason why the
result is unchanged even with such a small wave packet is that the  
neutrino is so light that its velocity $v_{\nu}$ is almost the
 light velocity. Consequently,  the two space time positions of the
 neutrino are almost on the light cone where the dominant contribution
in the amplitude  comes from, as it will be discussed in the next
section. In fact the
 neutrino of energy  $1~\text{GeV}/c^2$ and  the mass 
$1~\text{eV}/c^2$    has a velocity
\begin{eqnarray}
& &v/c=1-2\epsilon \\
& &\epsilon=({m_{\nu}c^2 \over E_{\nu}})^2=5\times 10^{-19}\nonumber,
\end{eqnarray} 
hence the neutrino propagates a distance 
\begin{eqnarray}
l=l_0(1-\epsilon)=l_0-\delta l,\delta l= l_0\times \delta,
\end{eqnarray}
when light propagates  the distance $l_0$. This difference of distance,
$\delta l$
becomes
\begin{eqnarray}
& &\delta l=5\times 10^{-17}~[\text{m}]; ~l_0=100~[\text{m}] \\ 
& &\delta l=5\times 10^{-16}~[\text{m}]; ~l_0=1000~[\text{m}] ,
\end{eqnarray}
which are much smaller than the sizes of the above wave packets
Eqs.~$(\ref{mean-free-pion})$ and $(\ref{mean-free-muon})$. Hence,
the neutrino amplitude at the nuclear target or the atom target should
show interference. The geometry of the neutrino interference is shown in Fig.~\ref{fig:geo} 

The electron neutrino interactions in detectors are 
\begin{eqnarray}
& &\nu_{e}+e^{-} \rightarrow e^{-}+\nu_{e} 
\label{nue-leptonic}\\ 
& &\nu_{e}+A \rightarrow e^{-}+(A+1) +X
\label{nue-hcharged}\\ 
& &\nu_{e}+A \rightarrow e  +A  +X.
\label{nue-hneutral}
\end{eqnarray}
The neutrino 
wave packet  $\sigma_{\nu}$ in processes $(\ref{nue-leptonic})$
is of order   $10^{-10}$~m and the neutrino 
wave packet  $\sigma_{\nu}$ in processes $(\ref{nue-hcharged})$ and
$(\ref{nue-hneutral})$ is of order   $10^{-15}$~m . They are treated in
the same way as the neutrino from the pion  decay.

Low energy neutrinos of order  MeV from the sun or reactors must be
treated separately and will be studied in a next paper.


\section{Neutrinos in hadronic collisions}
Applying the wave packet formalism,  we obtain the space-time dependent 
neutrino amplitude.  From this amplitude we study the transition
probability at finite time interval.   
\subsection{Semileptonic weak Hamiltonian and decay amplitude}

Semileptonic decay of pion is described by the weak Hamiltonian  
\begin{eqnarray}
& &H_{w}=g \int d{\vec x}{\partial_{\mu}
 }\phi(x)J_{V-A}^{\mu}(x)=-igm_{\mu}\int d{\vec x} \phi(x)J_5(x)\\
& &J_{V-A}^{\mu}(x)=\bar \mu(x)\gamma^{\mu}(1-\gamma_5)\nu(x),J_5(x)=\bar \mu(x)(1-\gamma_5)\nu(x),
\end{eqnarray}
where ${\phi(x)}$,~$\mu(x)$, and $\nu(x)$ are the pion field, muon
field, and neutrino field  without any ambiguity. In the above equations, 
$g$ is the coupling
strength, $\pi(x)$ , $J_{V-A}^{\mu}(x)$, and $J_5(x)$ are the pion field,
leptonic charged vector current, and leptonic pseudoscalar. 

\subsection{Neutrino production amplitude in hadronic collisions}
Neutrino production amplitude from an initial state
$|\alpha_i\rangle$ to an
final state $|\beta_f \rangle$ which includes the lepton pair is given by 
\begin{eqnarray}
T=\int d^4x \langle \beta_f |H_{w}(x)| \alpha_i \rangle.
\end{eqnarray}
When  the muon is chosen as  a plane wave  
and other states are wave packets specified
by central values of momenta, coordinates, and times,  
\begin{eqnarray}
& &|\alpha_i\rangle=   |{\vec p}_\text{proton};Nucleus  \rangle \\
& &|\beta_f \rangle=   |\mu,{\vec k};\nu,{\vec k}_{\nu},{\vec X}_{\nu},T_{\nu}  ;{\vec p}_{\pi},{\vec X}_{\pi},T_{\pi},hadrons        \rangle.     
\end{eqnarray}
Expressions of particle  states of the pion which ignores the life
time and of the neutrino and muon are expressed by 
one particle's matrix elements
\begin{eqnarray}
& &\langle 0|\phi(x)|{\vec p}_\text{pion},{\vec X}_\text{pion},T_\text{pion}  \rangle
\label{pion-wf}\\
& &=N_{\pi}\int d{\vec p}_{\pi} e^{-{\sigma_{\pi} \over 2}({\vec
 p}_{\pi}-{\vec p}_\text{pion})^2}e^{-iE({\vec
 p}_{\pi})(t-T_\text{pion}) + i{\vec
 p}_{\pi}\cdot({\vec x}-{\vec X}_\text{pion})}
\nonumber \\
& &\approx 
N_{\pi}e^{-{1 \over 2 \sigma_{\pi} }\left({\vec x}-{\vec X_{\pi}}-{\vec v}_{\pi}(t-T_{\pi})\right)^2}
e^{-iE({\vec p}_\text{pion})(t-T_\text{pion}) + i{\vec
 p}_\text{pion}\cdot({\vec x}-{\vec X}_\text{pion})} ,
\nonumber\\
& &\langle \mu,{\vec k};\nu,{\vec k}_{\nu},{\vec X}_{\nu},T_{\nu}|\bar
 \mu(x) \gamma_5 \nu(x) |0 \rangle\label{mu-nu-wf}\\
& & = \frac{N_{\nu}}{(2\pi)^{\frac{3}{2}}}\int d{\vec p}_{\nu}e^{-{\sigma_{\nu} \over 2}({\vec p}_{\nu}-{\vec k}_{\nu})^2}\left({m_{\mu}
 \over E({\vec p}_{\mu})}\right)^{1/2}\left({m_{\nu} \over E({\vec
 p}_{\nu})}\right)^{1/2}  \bar u({\vec p}_{\mu}) \gamma_5 \nu
 ({\vec p}_{\nu})\nonumber\\
& & \times e^{i\left(E({\vec p}_{\mu})t-{\vec
 p}_{\mu}\cdot{\vec x}\right)}e^{i\left(E({\vec p}_{\nu})(t-T_{\nu})-{\vec
 p}_{\nu}\cdot({\vec x}-{\vec X}_{\nu})\right)},  \nonumber \\
& & N_{\pi} = \left(\frac{\sigma_{\pi}}{\pi}\right)^{\frac{3}{4}},~N_{\nu} = \left(\frac{\sigma_{\nu}}{\pi}\right)^{\frac{3}{4}},\nonumber
\end{eqnarray}
where the spinor's normalization is
\begin{eqnarray}
\sum_{s} (u(p,s)\bar u(p,s))={\gamma p+m \over 2m}.
\end{eqnarray}

In the above equations, $(\ref{pion-wf})$ and $(\ref{mu-nu-wf})$, $\sigma_{\pi}$
,$\sigma_{\nu}$  are sizes of the pion wave packet of the
initial state and  of the neutrino wave packet of the final state. To
get the final expression  of Eq.~$(\ref{pion-wf})$, the pion momentum ${\vec p}$
is integrated and is replaced by the central value ${\vec p}_\text{pion}$. The
pion wave packet was 
estimated in the previous section and the neutrino wave packet size is
determined by the experimental apparatus. 
A neutrino interacts with a nucleus in the detector and the collision
with nucleus occurs incoherently. Hence the wave packet size of the neutrino
should be the nuclear size that is  the unit of detector. To study
neutrino interferences, we use the  nuclear size for $\sigma_{\nu}$. 

The
amplitude $T$ for  one pion to decay to the neutrino and muon  is written as,
\begin{eqnarray}
T &=& ig m_{\mu} N' \int d t d{\vec x} d{\vec p}_{\nu}
\langle\beta_f|\phi(x)|\alpha_i \rangle \left(\frac{m_{\nu}}{E(\vec{p}_{\nu})}\right)^{\frac{1}{2}}\label{amplitude}\\
& & \times e^{i\left(E(\vec{p}_{\mu})t-{\vec p}_{\mu}\cdot{\vec x}\right)}\bar{u}({\vec p}_{\mu}) \gamma_5 \nu({\vec p}_{\nu})e^{i\left(E({\vec p}_{\nu})(t-T_{\nu})-{\vec p}_{\nu}\cdot({\vec x}-{\vec X}_{\nu})\right) -{\sigma_{\nu} \over 2}({\vec p}_{\nu}-{\vec k}_{\nu})^2},\nonumber\\
N' &=& \frac{N_{\nu}}{(2\pi)^{\frac{3}{2}}} \left(\frac{m_{\mu}}{E(\vec{p}_{\mu})}\right)^{\frac{1}{2}},\nonumber\\
\langle \beta_f |\phi(x)|\alpha_i \rangle &=& \left(\frac{4\pi}{\sigma_{\pi}}\right)^{\frac{3}{4}}e^{-i\left(E({\vec
 p}_{pion})(t-T_{\pi})-{\vec p}_{pion}\cdot({\vec x}-{\vec
 X}_{\pi})\right)}\nonumber\\
& & \times e^{-{1 \over 2 \sigma_{\pi} }\left({\vec x}-{\vec X_{\pi}}-{\vec
 v}_{\pi}(t-T_{\pi})\right)^2}T_{\beta'_f,\alpha_i}.
\end{eqnarray}
In Eq.~$(\ref{amplitude})$, $(t,{\vec x})$ is the space-time 
coordinates where interaction takes place and the state $\beta_f'$ is
the hadron state in which one pion is taken away from the state $\beta_f$. From the dependence of the phase factor on the
momenta  of the muon, neutrino, and pion,  by using integration by part
it is shown that the average value of the energy  and momentum is
conserved.
The integration of the momenta multiplied by the integrand of Eq.~$(\ref{amplitude})$ satisfies 
\begin{eqnarray}
& &\langle p_\text{pion} \rangle=\langle p_{\mu}+p_{\nu} \rangle\\
& &\langle p_\text{pion}^2 \rangle=\langle (p_{\mu}+p_{\nu})^2 \rangle.
\label{average-momentum}
\end{eqnarray}
We use these relations later.

Due to the coordinate dependence of the wave
packets, the amplitude Eq.~$(\ref{amplitude})$ depends on the space time 
coordinates. Furthermore,  the integrand of Eq.~$(\ref{amplitude})$ 
depends upon the  space-time coordinate of the weak interaction where
the neutrino is produced and is given as 
  \begin{eqnarray}
T(t,{\vec x}) &=& ig m_{\mu} N'' \int d{\vec p}_{\nu}
\langle\beta_f|\phi(x)|\alpha_i \rangle 
 \times e^{i\left(E(\vec{p}_{\mu})t-{\vec p}_{\mu}\cdot{\vec
		x}\right)}\\
& &\bar{u}({\vec p}_{\mu}) \gamma_5 \nu({\vec p}_{\nu})e^{i\left(E({\vec p}_{\nu})(t-T_{\nu})-{\vec p}_{\nu}\cdot({\vec x}-{\vec X}_{\nu})\right) -{\sigma_{\nu} \over 2}({\vec p}_{\nu}-{\vec k}_{\nu})^2},
\end{eqnarray}
with a suitable normalization constant $N''$. This amplitude depends
upon the coordinates $(t,{\vec x})$ explicitly and is  not invariant under the 
translation.  So this satisfies peculiar properties of translational
non-invariant amplitude and the states of wide  momentum region play the
role. Even the infinite momentum states couple with
$\phi(x)$ and appears in $\beta_f$ and gives important contribution
to the probability at two different positions at finite times. This is
quite different
from the ordinary scattering amplitude where the infinite momentum
state decouples from the final state due to the energy 
momentum conservation.  We will study this point in detail later.
         

\subsection{Integration of neutrino momentum }
We compute the neutrino momentum integral of Eq.~$(\ref{amplitude})$ in
two methods. 
Gaussian integral is applied in the first one and stationary phase
approximation is applied in the second one.  The former method is valid 
in small time interval and the latter one is valid in large time
interval. In both methods, we obtain 
qualitatively same results. Especially the phase of neutrino wave
function has a particular form that is proportional to the square of
the mass and inversely proportional to the neutrino energy. 


\subsubsection{Gaussian integral}
For not so large $t-T_{\nu}$, the neutrino momentum ${\vec p}_{\nu}$
integration of Eq.~$(\ref{amplitude})$ is made by Gaussian integral around 
the central momentum ${\vec k}_{\nu}$. The amplitude becomes then,
 \begin{eqnarray}
T&=&igm_{\mu}\tilde N \int d t d{\vec x} \langle \beta_f|\phi(x)|\alpha_i \rangle
e^{i(E({\vec p}_{\mu})t-{\vec
 p}_{\mu}\cdot{\vec x})}\bar u({\vec p}_{\mu}) \gamma_5 \nu ({\vec k}_{\nu})
e^{i\phi}  \nonumber \\
& & \times \left({m_{\nu} \over E({\vec k}_{\nu})}\right)^{1/2} e^{-{1 \over 2\sigma_{\nu} }({\vec x}-{\vec X}_{\nu} -{\vec
v}_{\nu}(t-T_{\nu}))^2},
\end{eqnarray}
where $\tilde N$ is the normalization factor, $v_{\nu}^i$ is the  i-th 
component of the neutrino velocity, and $\phi$ is the phase of neutrino
wave function. They are   given by  
\begin{eqnarray}
& &\tilde N=\left(\frac{1}{2\pi}\right)^{\frac{3}{2}}\left(\frac{4\pi}{\sigma_{\nu}}\right)^{\frac{3}{4}}\left({m_{\mu}
 \over E({\vec p}_{\mu})}\right)^{1/2}\\
& &v_{\nu}^i={k^i_{\nu} \over E_{\nu}({\vec k}_{\nu})},~
\phi=E({\vec
 k}_{\nu})(t-T_{\nu})-{\vec  k}_{\nu}\cdot({\vec x}-{\vec X}_{\nu}).
\end{eqnarray}

The phase factor of the neutrino wave function, $\phi$, is rewritten by
substituting the central value ${\vec x}$ of neutrino's Gaussian
function  
\begin{eqnarray}
{\vec x}={\vec X}_{\nu}+{\vec v}_{\nu}(t-T_{\nu}).
\end{eqnarray}
We have the phase   
\begin{eqnarray}
\phi&=&E({\vec k}_{\nu})(t-T_{\nu})-{\vec k}_{\nu}\cdot({\vec x}-{\vec X}_{\nu})\label{phase}\\
&=&E({\vec k}_{\nu})(t-T_{\nu})-{\vec k}_{\nu}\cdot{\vec
 v}_{\nu}(t-T_{\nu})\nonumber\\
&=&{E^{~2}_{\nu}(\vec k_{\nu})-{\vec k}_{\nu}^2 \over E_{\nu}({\vec k}_{\nu})}(t-T_{\nu})= {{m_{\nu}}^2 \over E_{\nu}(\vec{k}_{\nu})} (t-T_{\nu}),\nonumber
\end{eqnarray}
which  has a typical form of the relativistic particle. The phase
becomes  proportional to the neutrino mass squared and inversely proportional to
the neutrino energy.     

\subsubsection{stationary phase approximation}
Integration on the neutrino momentum ${\vec p}_{\nu}$ in the amplitude 
at  macroscopic time difference $t-T_{\nu}$ is made by stationary
phase method. The stationary momentum is obtained from the stationary
condition,
\begin{eqnarray}
{\partial \over \partial p_{\nu}^i}\phi({\vec p}_{\nu})=0.
\label{stationary-phase}
\end{eqnarray}   
The equation for the stationary phase Eq.~$(\ref{stationary-phase})$ is 
written as,
\begin{eqnarray}
\left.{ p_{\nu}^i \over E(p_{\nu}) }\right|_{{\vec
 p}_{\nu}={\vec p}_{\nu}^{~(0)}}(t-T_{\nu})-(x-X)^i=0.
\end{eqnarray}
The square of the above equation leads
\begin{eqnarray}
& &\left.{ {{ \vec p}_{\nu}}^{~2} \over E^2(p_{\nu}) }\right|_{{\vec
 p}_{\nu}={\vec p}_{\nu}^{~(0)}}(t-T_{\nu})^2-({\vec x-\vec X})^2=0\\
& & 1+\left.{m^2 \over {{\vec p}_{\nu}}^{~2} }\right|_{{\vec
 p}_{\nu}={\vec p}_{\nu}^{~(0)}}={(t-T_{\nu})^2 \over ({\vec x-\vec X})^2 }.
\end{eqnarray}
Hence the phase factor at the stationary point is given by, 
\begin{eqnarray}
\phi({\vec p}_{\nu}^{~0})|_{{\vec
 p}_{\nu}={\vec p}_{\nu}^{~(0)}}&=&m\left((t-T_{\nu})^2-({\vec x}-{\vec
 X}_{\nu})^2\right)^{1/ 2}\\
&=&\left.{m^2 \over E_{\nu}({\vec p}_{\nu})}\right|_{{\vec
 p}_{\nu}={\vec p}_{\nu}^{~(0)}} (t-T_{\nu}).\nonumber
\end{eqnarray}
This phase agrees to Eq.~$(\ref{phase})$ obtained by the Gaussian
integral method.

Hereafter we use ${\vec p}_{\nu}$ instead ${\vec p}_{\nu}^{~(0)}$ and we
understand that the ${\vec p}_{\mu}$ in the following equations stands
for ${\vec p}_{\nu}^{~(0)}$ which is a function of time and space
coordinate. 
The decay amplitude becomes  
\begin{eqnarray}
T&=&i g m_{\mu} \tilde N \int d t d{\vec x} \langle \beta_f|\phi(x)|\alpha_i \rangle
e^{i\left(E({\vec p}_{\mu})t-{\vec
 p}_{\mu}\cdot{\vec x}\right)}
\bar u({\vec p}_{\mu}) \gamma_5 \nu ({{\vec p}_{\nu}})\nonumber \\
& & \times \left({m_{\nu} \over E({\vec k}_{\nu})}\right)^{1/2} e^{i {m_{\nu}^2 \over E_{\nu}} (t-T_{\nu})-{\sigma_{\nu} \over 2}\left({\vec
p}_{\nu}-{\vec k}_{\nu}\right)^2}, \label{amplitude2}\\
\tilde N &=& \frac{1}{(2\pi)^{\frac{3}{2}}}\left(\frac{4\pi}{\sigma_{\nu}}\right)^{\frac{3}{4}}\left({m_{\mu}
 \over E({\vec p}_{\mu})}\right)^{1/2}.
\end{eqnarray}

When the space time coordinates $( x_0,{\vec x})$ are
integrated, the delta function of the energy and momentum conservation
appears. The scattering amplitude  with this 
delta function has final states that have the same energy and momentum
with the initial states. On the other hand, the space and time dependent
amplitude $T(t,{\vec x})$ is not invariant under the translation and has
no delta function. So the  energy and momentum of the final state is not
necessary the same as the initial state. The states which do not satisfy 
the total energy and momentum conservation should be  included to get
consistent results from the completeness.  
This
scattering amplitude  shows the space and time dependent behavior, which
is a new information. 
So by
interchanging the order of the integration, we are able to obtain the
probability and other informations at the finite time interval.


\subsection{Decay probability and interference }
We find the probability of observing
neutrino at certain finite distance. Due to the finite distance effect
or finite time interval effect, the probability at the finite time
interval is not invariant 
under the translation. So special care is needed due to the non-standard
nature of the non-invariant probability. 
The integration over the momentum should be made with special care for
the translational non-invariant probability. The infinite momentum
states should be included 
 from the completeness of the physical space  unless  energy
 momentum conservation restrict. We will see that the  state of infinite
 momentum actually contribute in a manner that is almost equivalent to 
virtual state. These
 pseudo-virtual states
 of the infinite momentum are actually important for the probability of 
finite time to get the consistent result.  
We compute the  probability with   two different methods which are
applied  in previous subsection.   
\subsubsection{Gaussian integral}
Transition probability is a square of the above amplitude and is given
by
\begin{eqnarray}
|T|^2 &=& g^2 m_{\mu}^2 \left(\frac{4\pi}{\sigma_{\pi}}\right)^{\frac{3}{2}}|\tilde N|^2 \int d^4x_1 d^4x_2
\left|T_{\beta'_f,\alpha_i}\right|^2
S_{5}(s_1,s_2){m_{\nu} \over E({\vec k}_{\nu})}\nonumber\\
 &\times& e^{i {m_{\nu}^2 \over E_{\nu}} (t^1-T_{\nu})}e^{-i {m_{\nu}^2 \over E_{\nu}} (t^2-T_{\nu})}
e^{-{1 \over 2\sigma_{\nu} }\left({\vec x}^1-{\vec X}_{\nu} -{\vec v}_{\nu}(t^1-T_{\nu})\right)^2}e^{-{1 \over 2\sigma_{\nu} }\left({\vec x}^2-{\vec X}_{\nu} -{\vec v}_{\nu}(t^2-T_{\nu})\right)^2}
\nonumber \\
&\times& e^{-i\left(E({\vec p}_\text{pion})(t^1-T_{\pi})-{\vec p}_\text{pion}\cdot({\vec x}^1-{\vec
X}_{\pi})\right)}\times e^{i\left(E({\vec p}_\text{pion})(t^2-T_{\pi})-{\vec p}_\text{pion}\cdot({\vec x}^2-{\vec
X}_{\pi})\right)}\nonumber \\
&\times&e^{i\left(E({\vec p}_{\mu})t^1-{\vec p}_{\mu}\cdot{\vec x}^1\right)} \times
 e^{-i\left(E({\vec p}_{\mu})t^2-{\vec p}_{\mu}\cdot{\vec x}^2\right)}
\nonumber \\
&\times&e^{-{1 \over 2 \sigma_{\pi} }\left({\vec x}^1-{\vec X_{\pi}}-{\vec
 v}_{\pi}(t^1-T_{\pi})\right)^2}e^{-{1 \over 2 \sigma_{\pi} }\left({\vec x}^2-{\vec
 X_{\pi}}-{\vec v}_{\pi}(t^2-T_{\pi})\right)^2}, \label{probability}
\end{eqnarray}
where $S_{5}(s_1,s_2)$ stands for the products of Dirac
spinors and their  complex conjugates,   
\begin{eqnarray}
S_{5}(s_1,s_2)=\left(\bar u({\vec p}_{\mu})
 (1-\gamma_5) \nu ({{\vec p}_{\nu}})\right)\left(\bar u({\vec p}_{\mu})
 (1-\gamma_5) \nu ({{\vec p}_{\nu}})\right)^{*},
\label{spinor-1}
\end{eqnarray}
and its spin summation is  
\begin{eqnarray}
S^{5}&=&\sum_{s_1,s_2}S^{5}(s_1,s_2)\\
\label{spinor-2}
&=&{1 \over
 m_{\nu}m_{\mu}}2(p_{\mu}\cdot p_{\nu}).
\nonumber
\end{eqnarray}
We use the  relation Eq.~$(\ref{average-momentum})$ and write  $S_{5}$ as
\begin{eqnarray}
S^{5}={1 \over  m_{\nu}m_{\mu}} (m_\text{pion}^2- m_{\mu}^2).
\end{eqnarray}
\subsubsection{ Muon momentum integration}

When the muon in the final state is not observed, the muon momentum, ${\vec
p}_{\mu}$, is integrated. Since integral on coordinates $(t,{\vec x})$ is made
later, the energy-momentum conservation does not hold for $(t,{\vec x})$
dependent quantity. Hence the integration region of ${\vec p}_{\mu}$ is   whole
momentum region from the completeness of the state. The
muon of the infinite  momentum is produced as a real state but the time
dependent amplitude should include this momentum 
region. Otherwise Lorentz invariance does not hold  and meaningful
result is not obtained.

  Let the following function be $\Delta_{\mu}$ 
 \begin{eqnarray}
& &\Delta_{\mu} (\delta t,\delta {\vec x})= {\frac{1}{(2\pi)^3}}\int_{-\infty}^{\infty} {d {\vec p}_{\mu} \over E({\vec p}_{\mu})}e^{i\left(E({\vec
 p}_{\mu})\delta t-{\vec p}_{\mu}\cdot \delta {\vec x}\right)}\\
& &\delta t=t^1-t^2,\delta {\vec x}={\vec x}^1-{\vec x}^2,\nonumber
\end{eqnarray}
where time and space coordinates are fixed and so the four dimensional 
momentum ${p_{\mu}}$ is not restricted from the energy momentum
conservation. The integration region includes the infinite momentum
where the velocity becomes the light velocity.   Consequently this
function has a singular function near the light cone as 
\begin{eqnarray}
\Delta_{\mu}(\delta t,\delta {\vec x})=
{1 \over 4\pi}\delta(\lambda) \epsilon(\delta t) &-& i\frac{m_{\mu}}{8\pi
\sqrt{\lambda}}\theta(\lambda)\left[N_1\left(m_{\mu}\sqrt{\lambda}\right) -
			       i\epsilon(\delta
			       t)J_1\left(m_{\mu}\sqrt{\lambda}\right)\right] \nonumber\\
&+& i\theta(-\lambda)\frac{m_{\mu}}{4\pi^2 \sqrt{-\lambda}}K_1\left(m_{\mu}\sqrt{-\lambda}\right)\label{muon-correlation}
\\
& &\lambda=(\delta t)^2-(\delta {\vec x})^2,\\
\epsilon(\delta t) &=& \theta(\delta t) - \theta(-\delta t)
=\begin{cases}
 +1 \ \text{for} \ \ \delta t >0\\
 -1 \ \text{for} \ \ \delta t <0
 \end{cases}
\end{eqnarray}
and gives the most important contribution to certain  probability. We
investigate the effects of this most singular term later, which becomes 
most important in high energy limit and in  loop
amplitudes where the infinite momentum states give finite contribution. 
Actually $\Delta_{\mu}$ has several less singular terms and
oscillating terms. These terms  are not important  and we ignore in the
present work.

The probability for not so large $t-T_{\nu}$ is computed from the
amplitude Eq.~$(\ref{amplitude})$ and is given by
 \begin{eqnarray}
& &\int d{\vec p}_\text{muon} \sum_{s_1,s_2}|T|^2 \\
&= & g^2 m_{\mu}^2 |N_{\pi\nu}|^2\int d^4x_1 d^4x_2 |T_{\beta'_{f},\alpha_i}|^2{1 \over E_{\nu}}
e^{-{1 \over 2\sigma_{\nu} }\left({\vec x}^1-{\vec X}_{\nu} -{\vec
v}_0(t^1-T_{\nu})\right)^2}e^{-{1 \over 2\sigma_{\nu} }\left({\vec x}^2-{\vec X}_{\nu} -{\vec v}_0(t^2-T_{\nu})\right)^2}\nonumber \\
& & \times 
\Delta_{\mu}(\delta t,\delta {\vec x})
e^{i {m_{\nu}^2 \over E_{\nu}}\delta t} e^{-i\left(E({\vec p}_\text{pion})\delta t - {\vec p}_\text{pion}\cdot \delta
 \vec{x}\right)}
\times e^{-{1 \over 2 \sigma_{\pi} }\left({\vec x}^1-{\vec X_{\pi}}-{\vec
 v}_{\pi}(t^1-T_{\pi})\right)^2}\nonumber \\
& & \times e^{-{1 \over 2 \sigma_{\pi} }\left({\vec x}^2-{\vec
 X_{\pi}}-{\vec v}_{\pi}(t^2-T_{\pi})\right)^2} \nonumber \\
& &
N_{\pi\nu} =
\left(\frac{4\pi}{\sigma_{\pi}}\right)^{\frac{3}{4}}\left(\frac{4\pi}{\sigma_{\nu}}\right)^{\frac{3}{4}},~\delta
t =t_1-t_2,~\delta {\vec x}={\vec x}_1-{\vec x}_2.
\end{eqnarray}

\subsubsection{ Integration of the   interaction position}

Next we integrate the coordinates ${\vec x}_1$ and ${\vec x}_2$.
Integration on coordinates ${\vec x}_1$ and ${\vec x}_2$ are made by the
Gaussian integral on the neutrino wave packet and the
$\Delta_{\mu}(\delta t,\delta{\vec x})$,  
\begin{eqnarray}
& &I_0=\int d{\vec x}_1 d{\vec x}_2e^{-{1 \over 2\sigma_{\nu} }\left({\vec x}^1-{\vec X}_{\nu} -{\vec
v}_0(t^1-T_{\nu})\right)^2}e^{-{1 \over 2\sigma_{\nu} }\left({\vec x}^2-{\vec
X}_{\nu} -{\vec v}_0(t^2-T_{\nu})\right)^2}\Delta_{\mu}(\delta t,\delta {\vec
x}).
\nonumber \\
& &
\end{eqnarray}
By changing the variables ${\vec x}_1$ and ${\vec x}_2$ to ${\vec
X}={{\vec x}_1+{\vec x}_2 \over 2}$,~${\vec x}=\frac{{\vec x}_1-{\vec x}_2}{2}$, 
the above integration is made easily, and we have   
\begin{eqnarray}
I_0={2\pi \sigma_{\nu}\over 4\pi|t_1-t_2| }\epsilon( t_1-t_2).
\end{eqnarray}  
Finally the total probability is written as 
\begin{eqnarray}
& &\int d{\vec p}_\text{muon} \sum_{s_1,s_2}|T|^2 \\
&=&g^2 m_{\mu}^2 |N_{\pi\nu}|^2\int dt_1 dt_2 {1 \over E_{\nu}} 
{\sigma_{\nu} \over 2}
e^{-i\left(E({\vec p}_\text{pion}) {\delta t}-{\vec p}_\text{pion}\cdot{\delta {\vec
x}}\right)} \times {e^{i {m_{\nu}^2 \over
E_{\nu}} (t_1-t_2)} \over | t_1-t_2|}\epsilon(t_1-t_2)
\nonumber \\
& &\times e^{-{1 \over 2 \sigma_{\pi} }\left({\vec X}_{\nu}-{\vec X}_{\pi}+({\vec
 v}_{\nu}-{\vec v}_{\pi})(t^1-T_{\nu}) + \vec{v}_{\pi}(T_{\pi} - T_{\nu})\right)^2}
e^{-{1 \over 2 \sigma_{\pi} }\left({\vec X}_{\nu}-{\vec X}_{\pi}+({\vec
v}_{\nu}-{\vec v}_{\pi})(t^2-T_{\nu}) + \vec{v}_{\pi}(T_{\pi} - T_{\nu})\right)^2}. \nonumber
\end{eqnarray}
From the pion coherence length obtained in the previous section, the
pion Gaussian parts are regarded as
 \begin{eqnarray}
 e^{-\frac{1}{2\sigma_{\pi}}\left(\vec{X}_{\nu} - \vec{X}_{\pi} +
			     (\vec{v}_{\nu} - \vec{v}_{\pi})(t^1 -
			     T_{\nu}) + \vec{v}_{\pi}(T_{\pi} -
			     T_{\nu})\right)^2}\approx 1 \\
 e^{-\frac{1}{2\sigma_{\pi}}\left(\vec{X}_{\nu} - \vec{X}_{\pi} +
			     (\vec{v}_{\nu} - \vec{v}_{\pi})(t^2 -
			     T_{\nu}) + \vec{v}_{\pi}(T_{\pi} -
			     T_{\nu})\right)^2}\approx 1
 \end{eqnarray}
in a distance of our interest which is of order few $100$ m. In a larger
distance, this condition is not satisfied and the interference
disappears then.    
This condition is that the neutrinos produced in the different decay
area  overlap each other. Other situations where this condition is not
met, the interference  pattern becomes different.


\subsubsection{ Pion momentum integration}

\begin{figure}[t]
 \begin{minipage}{0.5\hsize}
  \centering
   \includegraphics[scale=.3]{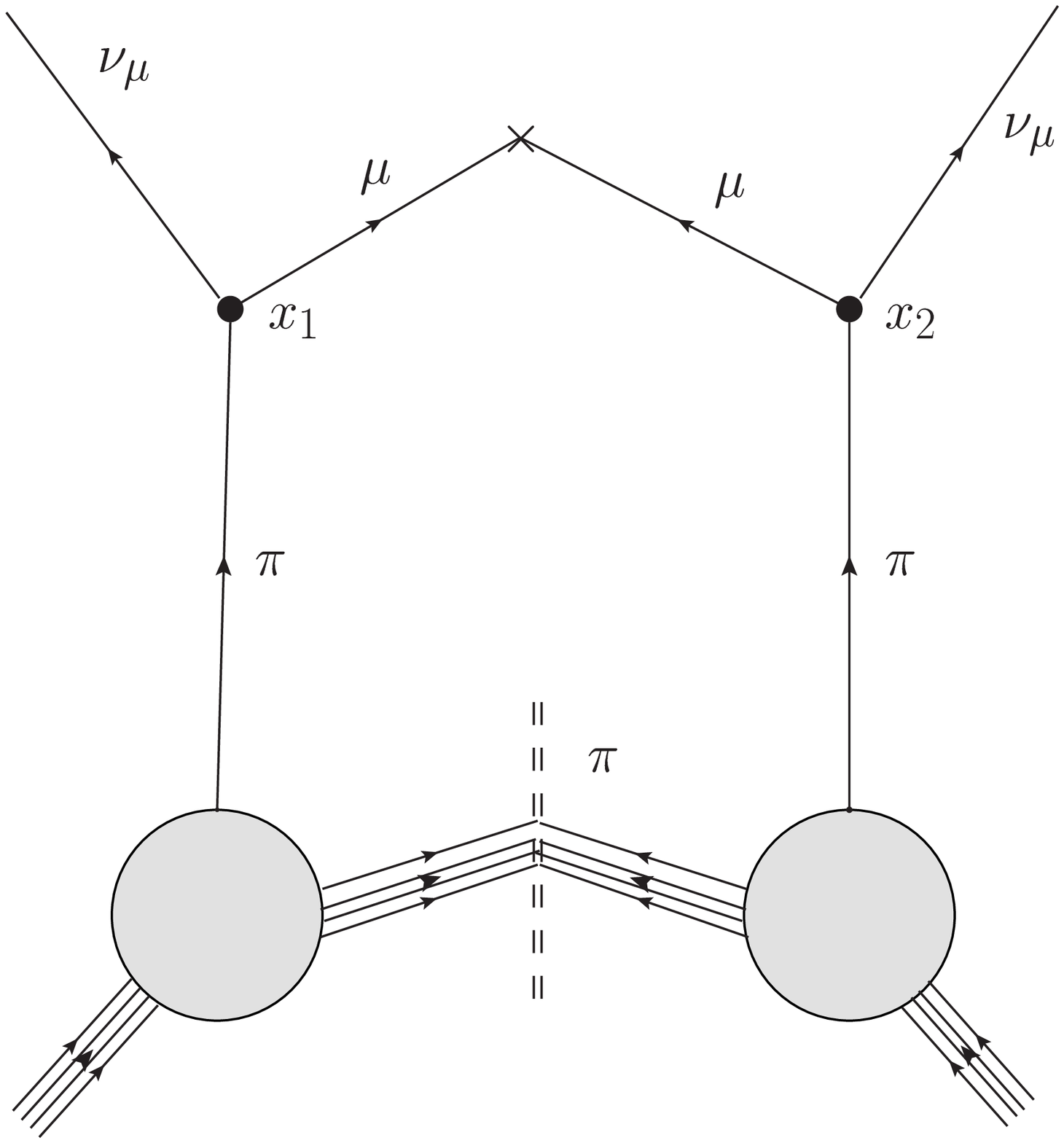}
\caption{Diagram of the normal term of transition probability
  for observing the
  neutrino produced at $x_1$ or $x_2$. All hadrons have
  finite energy
  and momentum. }
\label{fig:pi-standard}
 \end{minipage}\ \ \ \  
 \begin{minipage}{0.5\hsize}
  \centering
   \includegraphics[scale=.3]{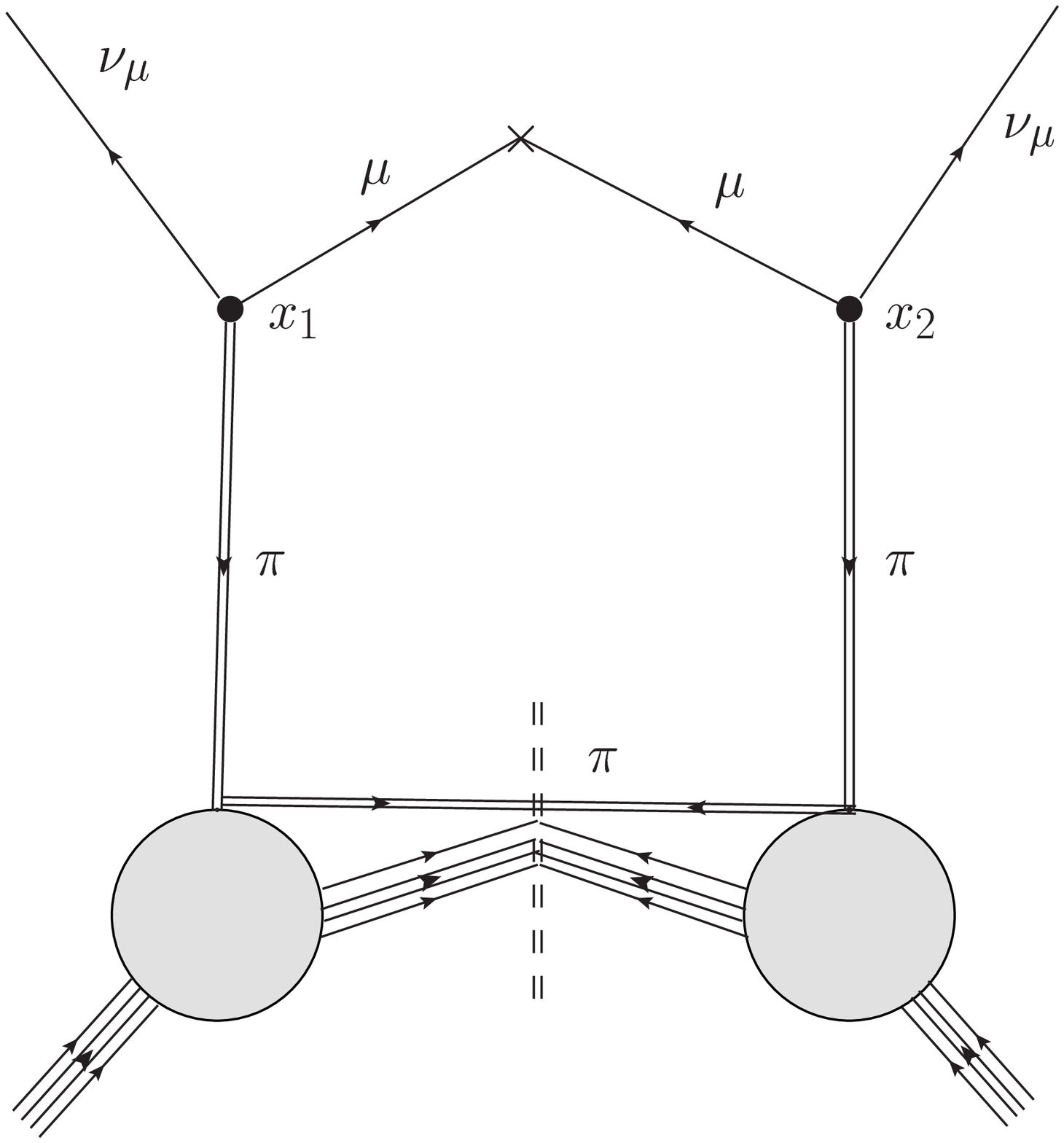}
\caption{Diagram of the anomalous  term of transition probability for observing the
  neutrino produced at $x_1$ or $x_2$. One pion along one line has the
  infinite  energy and momentum.}
\label{fig:pi-line}
\end{minipage}
\end{figure}

It is convenient to classify the diagram for the fixed values of $x_1$ and $x_2$ of the Eq.~$(\ref{probability} )$ into two
types. In the first one, the correlation function of two pions of
coordinates $x_1$ and $x_2$  is composed of those  of finite momenta of
on mass shell and we write this 
as  Fig.~\ref{fig:pi-standard}. In the second  one, the
pion correlation 
function is composed of   those  of infinite 
momentum and we write this as  Fig.~\ref{fig:pi-line}.  Because the 
coordinates $x_1$
and $x_2$ are fixed, momentum and energy flows at these points can have  
arbitrary values in these amplitude, which is a peculiar property of
position dependent amplitude.  The infinite momentum flow from the
${x_1}$ to the hadron part make some states in the  final state $\beta$ 
to have infinite momentum.  Hence the final state of the space-time dependent
amplitude, i.e., amplitude of fixed coordinates $x_1$ and $x_2$, should
include the infinite momentum sates from 
the completeness of the particle states. This infinite momentum states
contribute only to $(t,{\vec x})$ dependent quantities and to the
probability at the finite time. This states  do not
contribute to the standard scattering amplitude defined as the overlap
between the states at $t=\pm \infty$.  Hence this contribution is 
similar  to higher order quantum effect or one-loop effect where  virtual 
states of the
infinite momentum give a finite contribution. We will see that this term
gives the important
contribution to the interference term at  finite time. If the
integration over the space time $(t,{\vec x})$
is made first as in the ordinary scattering amplitude, the probability
has no dependence  on the space and time and these states of infinite 
momentum decouple from the integrated total amplitude at the infinite time.  

\begin{figure}[t]
 \centering
 \includegraphics[scale=.4]{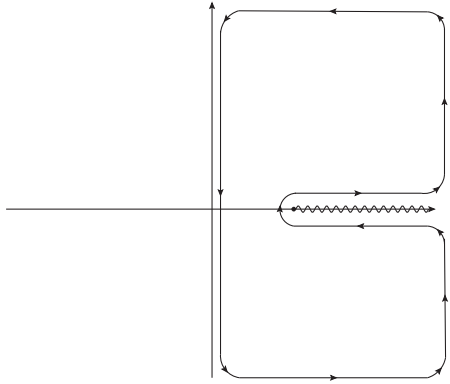}
 \caption{The energy contour in the integration along the real axis is
 changed into that along the imaginary axis by Wick rotation.}
\label{Wick-rotation}
\end{figure}
Pion in Fig.~\ref{fig:pi-standard} is constrained by the initial states
$\alpha$
and the final states $\beta$
of the hadronic part and has the maximum momentum allowed from the
energy and momentum conservation of the final state. The
pion in  Fig.~\ref{fig:pi-line}, however, is not constrained and can
have  infinite momentum.   Thus the two diagrams are different each
others. However
there is no way to separate directly two components using the
observables. In stead of a direct separation of two components, the
probability at the finite time is decomposed into two components, one is
the T-linear term and another is T-oscillating
term. Fig.~\ref{fig:pi-standard} contributes to the the T-linear term
and  Fig.~\ref{fig:pi-line} contributes to the both terms. Consequently
the T-oscillating term from  Fig.~\ref{fig:pi-line} is uniquely computed
from our formulation and we study this term.  The  pion 
momentum is taken in whole momentum region from the completeness of the 
states. The result gives a   Lorentz invariant term of the variable
$\lambda=(x_1-x_2)^2$. 

The pion correlation function of space time 
coordinates is obtained by integrating  the pion momentum and has two terms, 
\begin{eqnarray}
& &\Delta_\text{pion}(\delta t,\delta{\vec x})=\langle \beta_f |\phi(x_1)|\alpha_i
 \rangle \langle \alpha_i |\phi(x_2)^{\dagger}|\beta_f
 \rangle 
\label{pion-correlation}
\\
& &=F_\text{universal}(\lambda)+F_\text{nomral}({\delta t },{\delta {\vec
 x}}),\nonumber 
\end{eqnarray}
where the non-invariant term is given by 
\begin{eqnarray}
F_\text{normal}
=\sum_{{\vec p}_{\pi}} P({\vec p}_{\pi})e^{-i\left(E({\vec p}_\text{pion}) {\delta t^1}-{\vec p}_\text{pion}\cdot{\delta {\vec
x}^1}\right)}, 
\end{eqnarray}
where the momentum in the above equation is that of the original pion
momentum but not the momentum of the center of wave packet and is in the
finite range allowed by the energy and momentum conservation. This
correlation function is
a function of the combination ${\delta x}_{\mu} P_\text{initial}^{\mu}$ and   is
not invariant under Lorentz transformation of
${\delta x}$ only. 
The first one is  manifestly invariant under Lorentz transformation of
${\delta x}$ and a function of $\lambda=(\delta x)^2 $. 

First term in the right-hand side of Eq.~$(\ref{pion-correlation} )$ is
due to Fig.~\ref{fig:pi-line} where the intermediate
pion has the infinite momentum and the second term is due
to Fig.~\ref{fig:pi-standard} where the intermediate pion has finite momenta 
allowed from
the energy and momentum conservation. We argue further on 
the $\Delta_\text{pion}(\delta t,\delta {\vec x} )$ and
$\Delta_\text{muon}(\delta t,\delta {\vec x} )$  from operator product
expansions~\cite{Wilson-OPE} in \ref{App:OPE}. Particularly the
magnitude of~$F_\text{universal}(\lambda)$~is  estimated based on the Euclidean metric
integration which is obtained by deforming integration path of the
the intermediate energy along the real axis into the one along the
imaginary axis  by Wick rotation as in Fig.~\ref{Wick-rotation}. The integrand has a cut along the real
axis due to infrared divergence and the integral  thus defined  is Lorentz 
invariant and would be valid as a regularized value. We estimated the 
integral in the Euclidean metric and found a  finite value. We use this
value in this paper. 

The non-invariant part comes from real  pion decays where the pion
momentum has an upper bound.  The
invariant part comes from interference terms where the pion momentum
reaches infinity.  We see that the amplitude from the non-invariant term 
oscillates with time as 
\begin{eqnarray}
F_\text{normal} (\delta t,\delta{\vec x})=f_\text{normal} e^{i\omega_{\pi}  \delta t},
\label{normal-time}
\end{eqnarray}
of a frequency determined by the pion energy as
\begin{eqnarray}
\omega_{\pi}={m_{\pi}^2 \over E_{\pi}}.
\end{eqnarray} 
This frequency is a magnitude of a microscopic angular velocity that is 
much larger than ${m_{\nu}^2 \over E_{\nu}}$.

The invariant term at $\lambda=0$,   $F_\text{universal}(0)$  
gives a dominant contribution   at the light cone region. Hence, we
substitute  
\begin{eqnarray}
\Delta_\text{pion}(\delta t,\delta{\vec
 x})=F_\text{universal}(0)+f_\text{normal}e^{i\omega_{\pi} \delta t},
\label{pion-correlation2}
\end{eqnarray}
in the following calculations.


\subsection{Probability}
\subsubsection{Interference term}
Using the
correlation function  we have the final expression of the probability of
observing neutrino at the time T  or at the distance $L=c\text{T}$, 
\begin{eqnarray}
\int d{\vec p}_\text{muon} \sum_{s_1,s_2}|T|^2 
&=& N \int_0^{\text{T}} dt_1 dt_2  {e^{i {m_{\nu}^2 \over E_{\nu}}   (t_1-t_2) }
 \over (t_1-t_2)} \left(1 +
		   f_\text{normal}e^{i\omega_{\pi}(t_1-t_2)}\right)
\label{probability2}\\
&=&
\frac{\tilde{N}_{\text{prob}}}{E_{\nu}} ~\left[
					  F_\text{universal}(0)\times g(\text{T},\omega_{\nu}) + f_{\text{normal}}\times g(\text{T},\omega_{\pi}
	   + \omega_{\nu})\right]\nonumber
\end{eqnarray}
\begin{eqnarray}
\tilde N_{\text{prob}} &=& 
(2\sqrt{2}\pi)^3
\sigma_{\nu}
 f_{\mu},~
f_{\mu} = g^2m_{\mu}^2(m_{\pi}^2 - m_{\mu}^2),\nonumber \\
g(\text{T},\omega) &=& \text{T} \int^\text{T}_0 dX \frac{\sin \left(\omega X\right)}{X} +
  \frac{1}{\omega}\left\{\cos(\omega \text{T}) - 1\right\},\label{oscilating-term}\\
\omega_{\nu}&=&\frac{m_{\nu}^2}{E_{\nu}},~\omega_{\pi}=\frac{m_{\pi}^2}{E_{\pi}},~L=c\text{T},\nonumber
\end{eqnarray}
where L is the length of decay region. Eq.~$(\ref{probability2})$
depends on the neutrino wave packet size $\sigma_{\nu}$ and the pion
multiplicity but the number of neutrino events in the experiment is 
proportional to the initial energy, the neutrino reaction rate, the 
detector efficiency, and other parameters of the experiment  in addition 
to  Eq.~$(\ref{probability2})$.  We combine them to one constant $N_\text{exp}$, then
the event
number of neutrinos at the distance $L$ is 
proportional  to
\begin{eqnarray}
\frac{N_\text{exp}}{E_{\nu}}(F_\text{universal}(0)\times g(\text{T},\omega_{\nu}) + {f}_{\text{normal}}\times g(\text{T},\omega_{\pi}
	   + \omega_{\nu})).\label{probability-event}
\end{eqnarray} 

\begin{figure}[t]%
   \begin{center}
   \includegraphics[angle=-90,scale=.5]{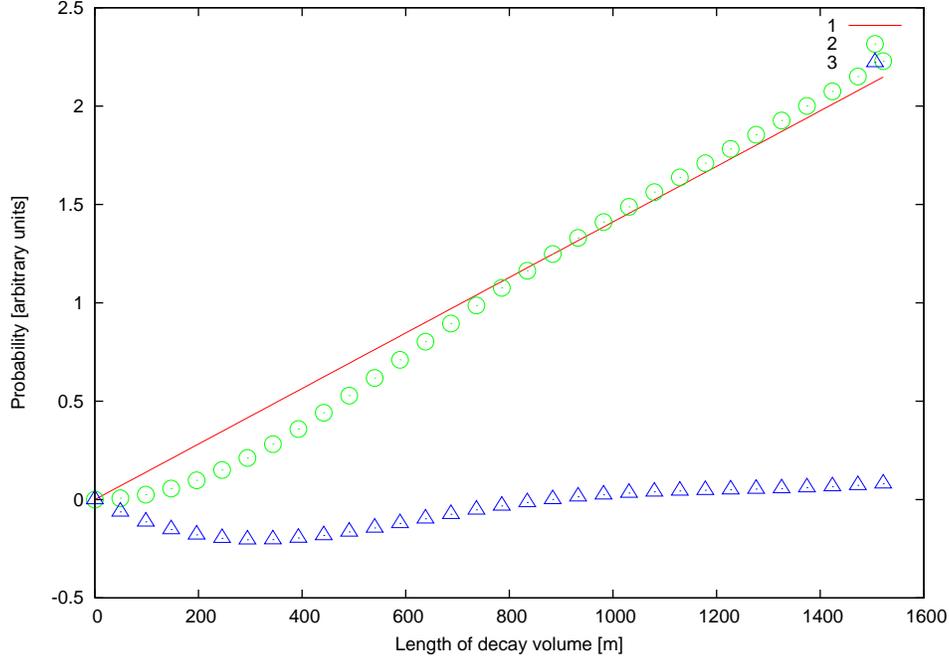}
   \end{center}
\caption{The universal term of the neutrino probability at a 
finite distance is given by circles and those that subtracted
the  T-linear term shown by a solid line  is given by triangles. 
The horizontal axis shows the distance in~[m] and the probability  is
  of arbitrary unit. The neutrino mass is $m_{\nu}=1~\text{eV}/c^2$}.
\label{fig:virtual-pi-total}
\end{figure}%

\begin{figure}[t]%
\begin{center}
  \includegraphics[angle=-90,scale=.5]{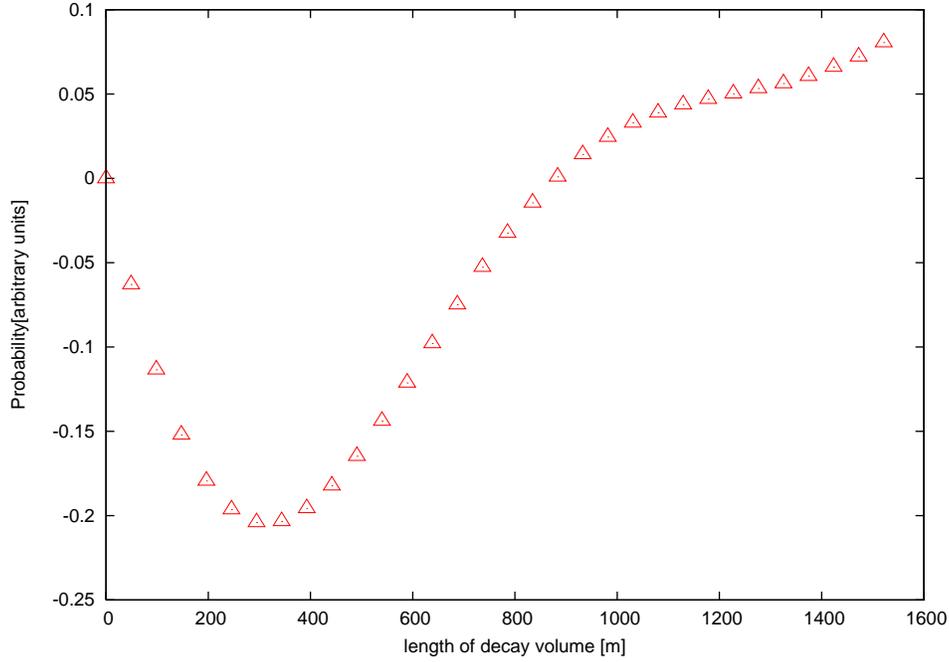}
   \end{center}
  \caption{After the T-linear term is subtracted, the probability shows 
clear oscillation.  The horizontal axis shows the distance in~[m] and the
  probability is of arbitrary unit.The neutrino mass is $m_{\nu}=1~\text{eV}/c^2$}.
 \label{fig:virtual-pi-sub}
\end{figure}

Eq.~$(\ref{oscilating-term})$  has
 a slowly oscillating term and
rapidly oscillating term. The latter term  oscillates so rapidly
that in the ordinary experiments only its average is seen. 
Hence, this term is regarded as a linear function of $T$,
\begin{eqnarray}
{f}_{\text{normal}}\times g(\text{T},\omega_{\pi}+\omega_{\nu}) =
{f}_{\text{normal}} \times \text{T}.
\end{eqnarray}
and we have the probability 
\begin{eqnarray}
\int d{\vec p}_\text{muon} \sum_{s_1,s_2}|T|^2 
=\frac{\tilde{N}_{\text{prob}}}{E_{\nu}} \left[F_\text{universal}(0) \times g(T,\omega_{\nu}) + {f}_{\text{normal}}\times \text{T} \right]\label{probability-2}.
\end{eqnarray}
The probability of observing the neutrino at the time $T$ is plotted in 
Fig.~\ref{fig:virtual-pi-total} for $m_{\nu}=1\text{eV}/c^2$. From this figure we see that it
consists of the $T$-linear term
and the small and slow oscillation term on top of the $T$-linear term.  
The $T$-linear term comes from the invariant term and also from the
non-invariant term.  So the relative magnitude of oscillation term is
not definite but from the calculation presented in \ref{App:OPE}, we expect
that the magnitude of the universal term is about the same as the the
non-universal term. Because the probability at the time $T$ has a linear term and the
universal slow oscillation term, the universal  term is extracted  by
subtracting the $T$-linear term from the total probability. The slowly
 oscillating  term thus obtained is given in
 Fig.~$\ref{fig:virtual-pi-sub}$. This shows the slow oscillation of
 showing small neutrino mass. 
Thus  we find the final form of the probability which 
has  the oscillating term with the frequency determined by the
neutrino's mass and the energy  and 
$T$-linear term
which comes from the Lorentz non-invariant term 
in Eq.~$(\ref{pion-correlation})$.  
\begin{figure}[t]
   \begin{center}
   \includegraphics[angle=-90,scale=.5]{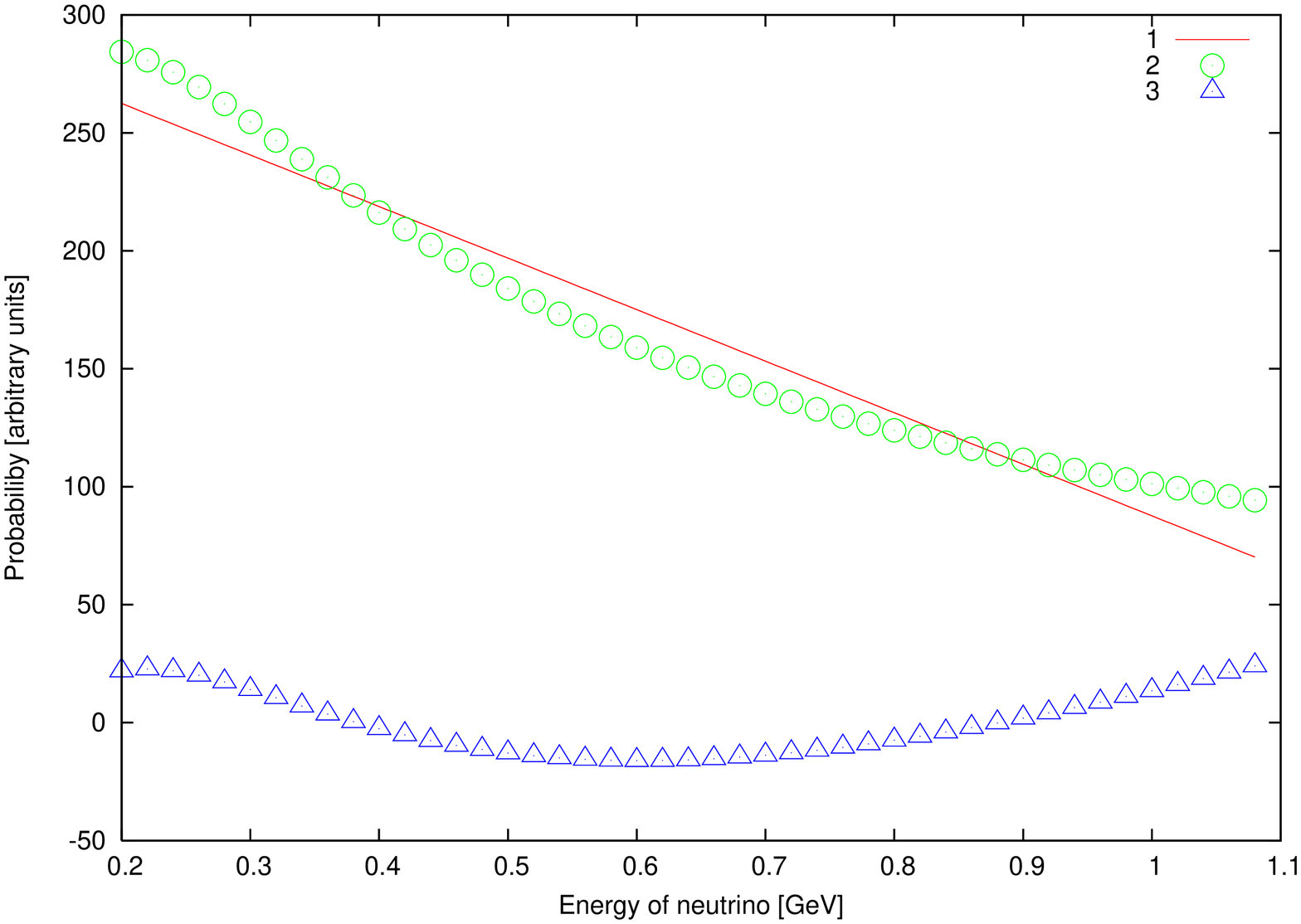}
   \end{center}
\caption{ The energy dependence of the neutrino probability is plotted.
 The probability at $T$ is given by circles and those that subtracted
the  T-linear term shown by the solid line  is given by triangles. The
 horizontal axis shows the neutrino energy in [GeV] and the vertical
 axis shows the neutrino probability in the arbitrary unit.The neutrino
 mass is $m_{\nu}=1~\text{eV}/c^2$. The distance is $200$~m}
\label{fig:virtual-pi-totalE}
\end{figure}%
 
 \begin{figure}[t]
\begin{center}
  \includegraphics[angle=-90,scale=.5]{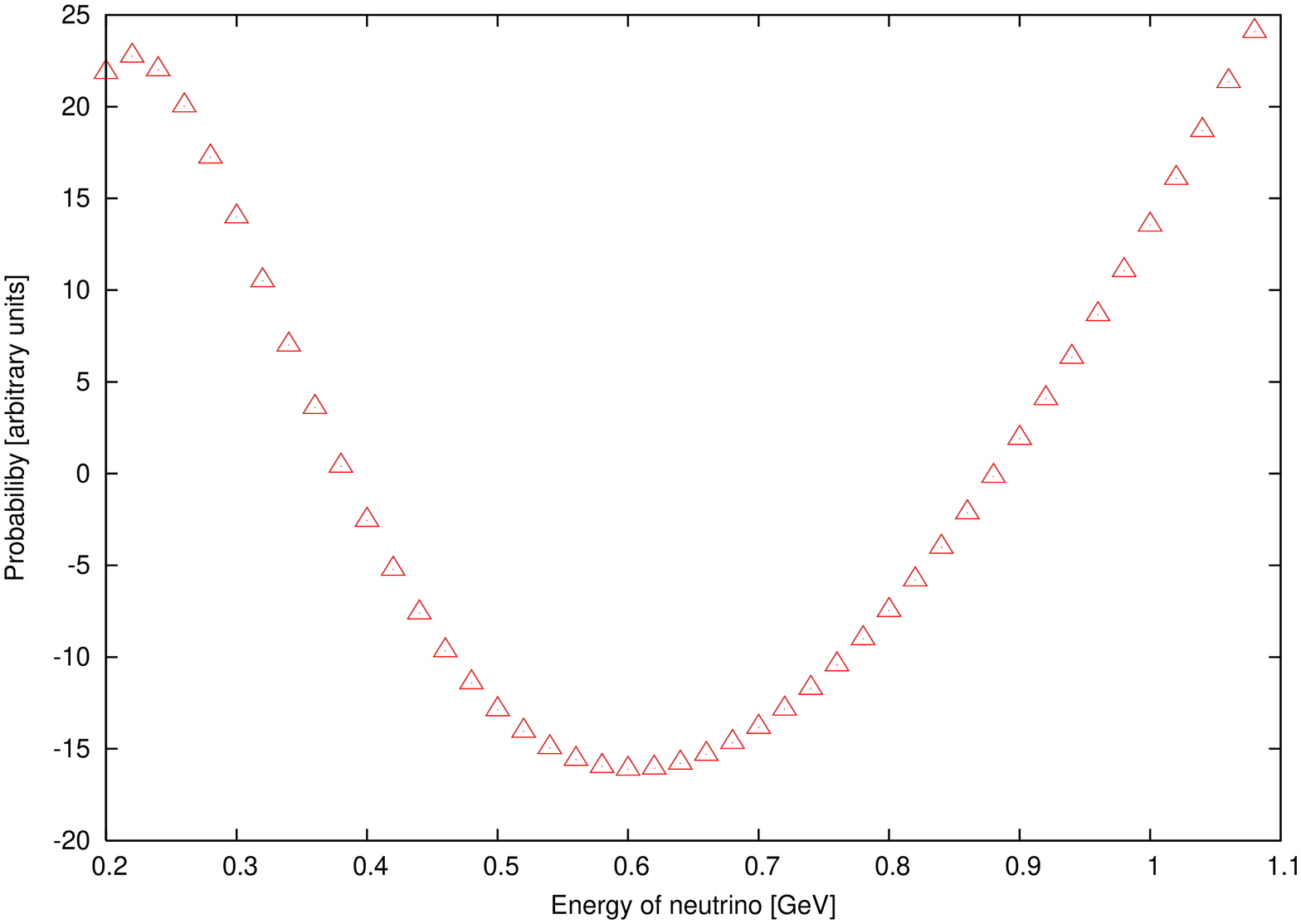}
   \end{center}
  \caption{The energy dependence of the oscillating neutrino probability
  that is subtracted the background term from the total probability is
  plotted. The horizontal axis shows the neutrino energy in [GeV] and
  the vertical
 axis shows the probability in the arbitrary unit.The neutrino mass is
  $m_{\nu}=1~\text{eV}/c^2$. The distance is $200$~m}
 \label{fig:virtual-pi-subE}
\end{figure}%

The slowly 
oscillating term is determined  by the Lorentz invariant term in
Eq.~$(\ref{pion-correlation})$.   A typical oscillation length $L_0$ is 
\begin{eqnarray}
L_0~[\text{m}] ={E_{\nu} \hbar c \over m_{\nu}^2 }= 200{E_{\nu}[\text{GeV}/c^2] \over
 m_{\nu}[\text{eV}/c^2]^2 }.
\end{eqnarray}
This length is quite interesting value for the observational point. We
use $m_{\nu}=1~\text{eV}/c^2$ throughout this paper.

The oscillating probability depends upon the energy $E_{\nu}$ and time
$T$ and its signature is observed also in the energy dependence of the
probability at a fixed $T$.  This 
total probability multiplied by the neutrino energy is plotted in Fig.~$\ref{fig:virtual-pi-totalE}$ and the
the probability obtained by subtracting the smooth background from the
total probability is given in Fig.~$ \ref{fig:virtual-pi-subE}$. The slow
oscillation due to the universal term is seen clearly.

\subsubsection{muon in pion decays}
When the muon is observed in the same processes, the anomalous oscillation
is determined by the muon mass and energy as ${m_{\mu}^2 \over E_{\mu}}$
and the muon mass is larger than the neutrino mass by $10^8$. Hence the
oscillation length is smaller than that of the neutrino by $10^{16}$. For
the muon of energy one GeV/c, the oscillation length is order $10^{-12}$~m.
This value is too small to observe  in experiments.  

Since it is hard to see 
anomalous oscillation of this length, it is meaningful  to study an average 
probability. The total muon flux 
includes the effect of the anomalous term and 
is written as 
\begin{eqnarray}
& &F_\text{muon}=F_{1n}+F_{1a}\\
& &F_{1n}= C\langle n_{\pi}\rangle \\
& &F_{1a}= C'\langle n_{\pi}\rangle  ,
\end{eqnarray}
where $F_{1n}$ is due to naive decay of pions without interference
effect and so is proportional to pion number, whereas  $F_{1a}$ is due
to the interference term that gives the anomalous oscillation of the
neutrino. Because we ignored the non-singular term in Eq.~$(\ref{muon-correlation})$, it is
hard to know the magnitude of rapid oscillation term and the total flux.  
We expect that
the $F_{1a}$ has a same magnitude as $F_{1n}$ from the calculation of
\ref{App:OPE}. 
\subsubsection{Stationary phase approximation}
So far the calculation is based on the Gaussian integration on the
neutrino momentum. The result is the same in another calculational
method of using the stationary phase approximation.

The transition probability is a square of the Eq.~\ref{amplitude2} and is given
by
\begin{eqnarray}
|T|^2 &=& \left(\frac{4\pi}{\sigma_{\pi}}\right)^{\frac{3}{2}}|\tilde{N}|^2\int
 d^4x_1 d^4x_2 
\left|T_{{\beta}^{~\prime}_f,\alpha_i}\right|^2
S_{5}(s_1,s_2)\left({m_{\nu} \over E^1_{\nu}E^2_{\nu}}\right)^{\frac{1}{2}}\nonumber\\
 &\times& e^{i {m_{\nu}^2 \over E_{\nu}} (t^1-T_{\nu})}e^{-i {m_{\nu}^2
  \over E_{\nu}} (t^2-T_{\nu})}
e^{-{\sigma_{\nu}
\over 2}\left( ({\vec p}_{\nu}^1-{\vec k}_{\nu})^2+ ({\vec p}_{\nu}^2-{\vec k}_{\nu})^2\right)}
\nonumber \\
 &\times& e^{-i\left(E({\vec p}_\text{pion})(t^1-T_{\pi})-{\vec p}_\text{pion}\cdot({\vec x}^1-{\vec
X}_{\pi})\right)}\times e^{i\left(E({\vec p}_\text{pion})(t^2-T_{\pi})-{\vec p}_\text{pion}\cdot({\vec x}^2-{\vec
X}_{\pi})\right)}\nonumber \\
&\times&e^{i\left(E({\vec p}_{\mu})t^1-{\vec p}_{\mu}\cdot{\vec x}^1\right)} \times
 e^{-i\left(E({\vec p}_{\mu})t^2-{\vec p}_{\mu}\cdot{\vec x}^2\right)}\nonumber \\
&\times&e^{-{1 \over 2 \sigma_{\pi} }\left({\vec x}^1-{\vec X_{\pi}}-{\vec
 v}_{\pi}(t^1-T_{\pi})\right)^2}e^{-{1 \over 2 \sigma_{\pi} }\left({\vec x}^2-{\vec
 X_{\pi}}-{\vec v}_{\pi}(t^2-T_{\pi})\right)^2}, \label{probability-stationary-phase}
\end{eqnarray}
where $S_{5}(s_1,s_2)$ stands for the products of Dirac
spinors and their  complex conjugates defined in Eq.~$(\ref{spinor-1})$  and its spin 
summation of Eq.~$(\ref{spinor-2})$.
The products of $S_5$ with pion momentum is computed
similarly 
and the probability for the unseen muon in which muon momentum is integrated 
is
\begin{eqnarray}
& &\int d{\vec p}_{muon} \sum_{s_1,s_2}|T|^2 \\
&=&|N_{\pi\nu}|^2\int d^4x_1 d^4x_2
 \left|T_{{\beta}^{~\prime}_f,\alpha_i}\right|^2
\left({1 \over E_{\nu}^1E_{\nu}^2}\right)^{\frac{1}{2}} e^{-{\sigma_{\nu} \over 2
}\left(({\vec p}_{\nu}^1-{\vec k}_{\nu})^2 +({\vec p}_{\nu}^2-{\vec
k}_{\nu})^2\right)} 
e^{-i\left(E({\vec p}_\text{pion}) {\delta t}-{\vec p}_\text{pion}\cdot{\delta {\vec
x}}\right)} \nonumber \\
& &\times \Delta(\delta t,\delta {\vec x})
e^{i {m_{\nu}^2 \over E_{\nu}^1} (t^1-T_{\nu})}e^{-i {m_{\nu}^2 \over
  E_{\nu}^{2}  } (t^2-T_{\nu})}e^{-{1 \over 2 \sigma_{\pi} }\left({\vec x}^1-{\vec X_{\pi}}-{\vec v}_{\pi}(t^1-T_{\pi})\right)^2}e^{-{1 \over 2 \sigma_{\pi} }\left({\vec x}^2-{\vec X_{\pi}}-{\vec v}_{\pi}(t^2-T_{\pi})\right)^2}. \nonumber
\end{eqnarray}
The final expression  has a simple form and is the same as the Gaussian
integral methods.
\section{Neutrino in real  pion decay}
So far the initial state are the proton and nucleus and the pion is
unidentified and is equivalent to intermediate state.

We study the case where the pion is identified and is in the initial
state of the scattering.  Let the momentum of the pion is $q_\text{pion}$,
then the decay amplitude of the pion is given as 
\begin{eqnarray}
& &T=gm_{\mu}\int d^4x ~\langle 0|\phi(x)|q_\text{pion}\rangle  \langle p_{\mu},\nu(x)| J_5(x)|0\rangle\\
& &=gm_{\mu}\int d^4x ~e^{-i(q_\text{pion}-p_{\nu})\cdot x} \langle p_{\mu}| \mu(x) |0 \rangle
 (1-\gamma_5)u({\vec p}_{\nu})\nonumber 
\end{eqnarray} 
After the integration of the momentum, the total probability becomes
\begin{eqnarray}
& &\int d{\vec p}_\text{muon} |T|^2 \\
& &=g^2 m_{\mu}^2\int d^4{x_1} d^4{x_2} 
~e^{-i(q_\text{pion}-p_{\nu})\cdot (x_1-x_2)}\Delta_{\mu}(x_1-x_2)S_5\nonumber
\end{eqnarray}
In the above result, the oscillation is so rapid that can not be
observed and the time average is found as 
\begin{eqnarray}
& &\int  d{\vec p}_\text{muon} |T|^2\\
& &=g^2 m_{\mu}^2(m_{\pi}^2-m_{\mu}^2)\int d^4{x_1} d^4{x_2}~ 
e^{-i(q_\text{pion}-p_{\nu})\cdot (x_1-x_2)}\Delta_{\mu}(x_1-x_2)\nonumber 
\end{eqnarray}
The above neutrino probability has the same property as the normal term 
Eq.~$(\ref{normal-time})$ and 
shows no long distance
interference. 

In higher order effect, the momentum of the intermediate
state includes infinitely large value, so the light cone behavior is
modified. Consequently it is expected that the long distance
interference is generated. 
\begin{figure}[t]
  \centering
   \includegraphics[angle=-90,scale=.5]{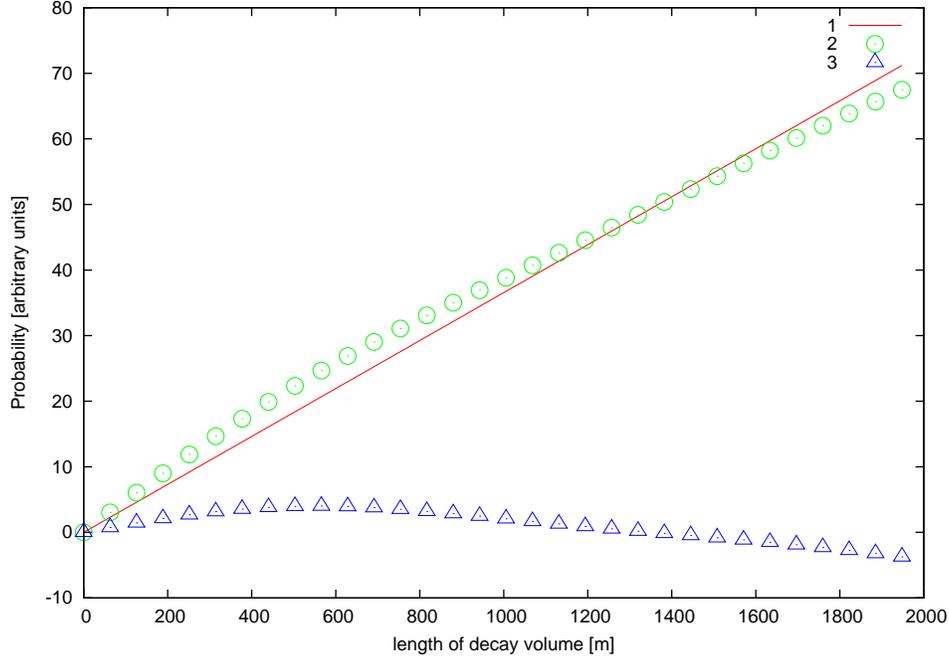}
  \caption{The  neutrino probability of the
 anomalous term is shown by circles and those that subtracted
the  T-linear term (solid line)  is shown by tri-angles. 
The horizontal axis shows the distance in~[m] and the magnitude is
  of arbitrary unit.The small oscillation
 is seen.The neutrino mass is $m_{\nu}=1\text{eV}/c^2$.}
  \label{fig:real-pi-toral}
 \end{figure}
 \begin{figure}[t]
  \centering
   \includegraphics[angle=-90,scale=.5]{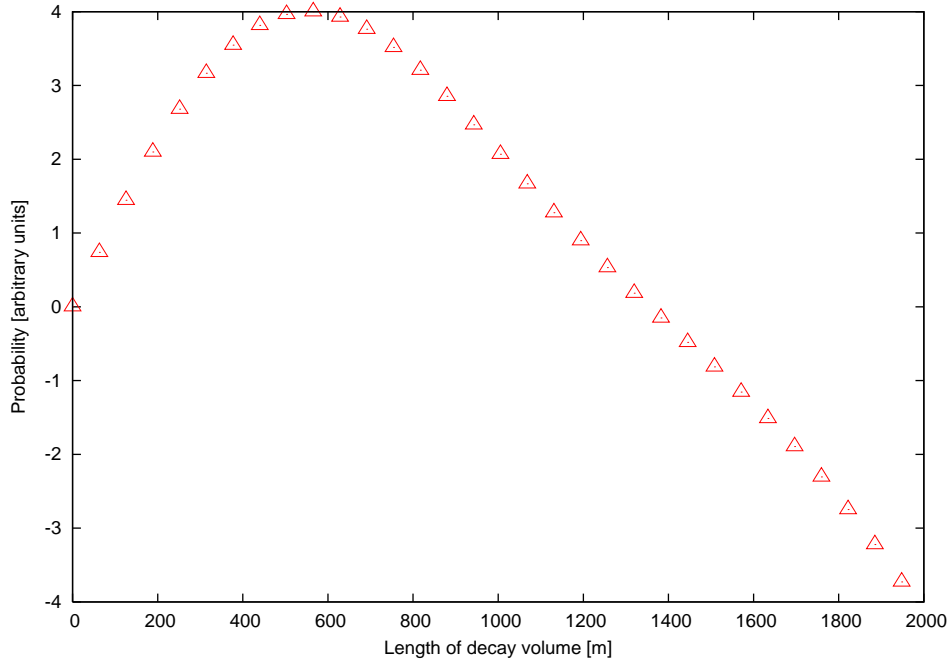}
  \caption{The length  dependence of the oscillating neutrino probability
  that is obtained subtracting the T-linear term from the total
  probability.The horizontal axis shows the distance in~[m] and the magnitude is
  of arbitrary unit.The neutrino mass is $m_{\nu}=1\text{eV}/c^2$.}
\label{fig:real-pi-sub}
\end{figure}%
The lowest higher order correction is given by the pion exchange term.
 In this case the external pion momentum is
included in the momentum of the intermediate pion and its effect remains
in  the final result,
\begin{eqnarray}
 & &\int  d{\vec p}_\text{muon} |T|^2\\
& &=g^2 g_{\pi\pi}^2 \int d^4{x_1} d^4{x_2} 
\Delta_{\mu}(x_1-x_2)\Delta_{\pi}({ p}_\text{pion},x_1-x_2)S_5
e^{ip_{\nu}\cdot (x_1-x_2)},\nonumber 
\end{eqnarray}
where the correlation function $\Delta_{\pi}({ p}_\text{pion},x_1-x_2) $
depends on the external pion momentum and is not a function of
$\lambda$. $\Delta_{\pi}({ p}_\text{pion},x_1-x_2) $ is computed in the
\ref{App:OPE} and becomes 
\begin{eqnarray}
& &\Delta_{\pi}({ p}_\text{pion},x_1-x_2)\\
& &={1 \over p_\text{pion}(x_1-x_2)} {1 \over 64\pi^2}{1 \over m^2}.\nonumber
\end{eqnarray}
   
The probability of observing the neutrino in the real pion decay is
given in Fig.~$(\ref{fig:real-pi-toral})$, which is composed of the $T$
linear term and the oscillating term. Because it is hard to know the
relative size of the one loop amplitude with the tree amplitude, the
relative magnitude of the oscillation term is arbitrary. It is known
that the non-universal terms from the tree level and the one-loop level 
behaves like the $T$-linear term and the universal term
oscillates. Hence the oscillating term is obtained by subtracting the
$T$-linear term from the total probability. The result is plotted in       
Fig.~$(\ref{fig:real-pi-sub})$ as a slow oscillation.
\section{Neutrinos from muon decay}
\subsection{Leptonic weak Hamiltonian and three  body decay amplitude}
Muon decays to an electron and two neutrinos. One is  an electron
neutrino and the other is an muon neutrino. We study the flux of each 
neutrino at a certain distance $L$ in high energy region.

Leptonic decay of muon is described by the leptonic weak Hamiltonian 
\begin{eqnarray}
& &H_{w}={G_{F} \over \sqrt 2 }\int d{\vec x}{J_{V-A}}_{\alpha} {J_{V-A}^{\alpha}}^{\dagger}(x)\\
& &J_{V-A}^{\alpha}(x)=\bar
 \mu(x)\gamma^{\alpha}(1-\gamma_5)\nu_{\mu}(x)+
\bar e(x)\gamma^{\alpha}(1-\gamma_5)\nu_e(x)
\end{eqnarray}
without any ambiguity. In the above equations, $G_F$ is the Fermi coupling
constant, $\mu(x)$ is the muon field,  $e(x)$ is the electron field,
$\nu_{\mu}(x)$ is the muon neutrino field, and $\nu_{e}(x)$ is the
electron neutrino field. $J_{V-A}^{\alpha}(x)$ is the leptonic
charged current. 

\subsection{Neutrinos from muon three body decay}
Muon decay is treated in a similar manner as the previous pion
 decay. Here we assume that the muon correlation function has two
 components. First one  is a function of $\lambda={\delta x}^2$ and is invariant
 under Lorentz transformation of the coordinates ${\delta x}$. Second one is
a function of the product ${\delta x}_{\mu}{P_\text{muon}}^{\mu}$ and has the
 same property as the normal term of the pion decay. 
 
 So
 the decay amplitude and  decay probability are computed in Gaussian 
integral method. We integrate unseen particle's momenta and average over
 the initial muon momentum. 

Leading term of expectation value of the
 weak leptonic currents  at the light cone is 
\begin{eqnarray}
& &\langle 0|J_{\mu_1}(x_1){J^{\mu_2}(x_2)}^{\dagger}|0\rangle
\label{two-lepton-spectrum}\\
& &=8(-{ \partial}^2 g_{\mu_1\mu_2} +{\partial_{\mu_1}
 \partial_{\mu_2}}) i\epsilon({(x_1^0-x_2^0)}) \delta'(\lambda)\nonumber.
\end{eqnarray}
Because the electron and one unseen neutrino are not observed and
their momenta are integrated in the infinite regions and the above
correlation $(\ref{two-lepton-spectrum})$ is more singular than the
previous one
particle case.  

The normal term gives the   rapid oscillation term and its time average
is proportional to the time $T$.   The one loop term is generated by QED
correction.
As is discussed in the \ref{App:OPE}, the fully invariant term does not
exist  in QED one loop correction but the non-oscillating power
correction term which  becomes roughly $\alpha$ times the magnitude of
the normal term,
\begin{eqnarray}
f_\text{universal}=\alpha {1 \over p_{\mu}(x_1-x_2)}\times  |f_\text{normal}|
\end{eqnarray}
do exist. This term gives an oscillation of the frequency determined by
the neutrino mass and energy.  

 Finally we have the similar expression for the decay probability 
as the previous pion decay,    
\begin{eqnarray}
& &\int d{\vec p}_\text{electron} d{\vec p}_{\nu'}\sum_{s_1,s_2}|T|^2 
=\tilde N_{\mu} \{g_1(T,\omega_{\nu}) + g_2(T,\omega_{\nu})\}+\tilde f_{\mu,\text{normal}}, \label{probability-3}\\
& & g_1(T,\omega) = T\int^T_0 dX \frac{\sin(\omega X)}{X^2},~g_2(T,\omega) = \int^T_0 dX \frac{\sin(\omega X)}{X}.\nonumber
\end{eqnarray}
Although the oscillating term that is due to the invariant term is much 
smaller than the ordinary term, the observation of this term may be
possible.

In muon decay one muon neutrino and one electron  neutrino are produced.
They are linear combination of three mass eigenstates and a unitary MNS
matrix combines flavour eigenstates with mass eigenstates. 
\subsubsection{Electron from muon decay}
If the electron is measured and two neutrinos are unseen we have the
probability of the electron at the distance $L$,
\begin{eqnarray}
& &\int d{\vec p}_{\nu} d{\vec p}_{\nu'}\sum_{s_1,s_2}|T|^2 
\label{probability-4}
=\tilde N_{e} \{g_1(T,\omega_{e}) + g_2(T,\omega_{e})\}+\tilde f_{e,\text{normal}},\\
& & \omega_e = \frac{m_e^2}{E_e},
\end{eqnarray}
where the oscillation length is given by 
\begin{eqnarray}
L_0^e= {E_e \hbar c \over m_e^2}.
\end{eqnarray} 
The oscillation length of the electron is given by
\begin{eqnarray}
L_0^e=  10^{-8}[\text{m}] \times ({E_e/\text{GeV}})
\end{eqnarray}
and is too small if the energy is a few  GeV. In the ultra-relativistic
energy region, this oscillation may become observable. 



\section{Summary and implications}

 In this paper, we  
showed  that one particle states  of decaying particles that produce 
neutrinos are described using  wave packets of finite
coherence lengths and studied its implications to the neutrino
interferences. 
 
The wave packet size was  determined
either from particle production processes and detection  processes.
In the former, a finite mean free path in matter is the origin of the wave
packet. The finite mean free path makes one particle to have a finite
spatial extension and
a finite momentum uncertainty. The state of a finite mean free path 
is a non-stationary  state and is varied with  time and space.  In the
latter, a finite size of the unit of detector is the origin of the 
wave packets. The wave packet sizes of the proton,  pion, muon, and the neutrino
were estimated and were used in analyzing high energy neutrino reactions.  

 Since the overall phase of wave packet during propagation is determined
 by the time
 component that is proportional to the energy  and the 
space component that is proportional to the momentum and the space
 position and the time position are connected each others, both effects
 are taken into account simultaneously. Due to the
relativistic invariance in the energy and momentum and in the time and
space position,  both  terms in the total  phase are almost 
cancelled and the total phase becomes small number that is proportional
 to the mass squared and inversely proportional to the energy. Consequently
when the neutrino is described by the wave packet, the space-time
dependent probability of the neutrino is found and the above overall
 phase or its difference becomes  observable.
We showed that the time dependent interference of the neutrino in the 
processes of the decays of pion or muon reveals this phase.

 The time dependent probability of observing the neutrino at finite distance
was  calculated for high energy collisions  and the anomalous oscillating
term was found. This term
has the origin in  the higher order
quantum effects where the infinite momentum virtual states play
the important role. 
The new universal term is  manifestly 
invariant under the Lorentz transformation of the coordinates and gives
the   most important contribution in the operator product near the
light cone region. 
Because the neutrino's velocity is almost the light velocity, the
time dependent probabily of finding the neutrino is determined by this
universal term. The probabily of finding the
neutrino at finite medium time  is oscillating  with the slow angular 
velocity in  Eq.~$(\ref{probability-2})$.  Since the angular velocity is 
determined by the neutrino mass and energy, the absolute value of the 
neutrino mass would be found from the neutrino interference oscillations.

Due to the
relativistic invariance, the correlation function $\Delta_{\mu}$ and
others become  functions of the
Lorentz invariant combination $\lambda$. The space-time points that satisfies
$\lambda=0$ are on the light-cone surface and  infinite
number of points are on the surface. This is a feature of a
relativistic invariant system and is a reason why the interference of
the present work occurs. 
For a non-relativistic system, in a stationary state of the same
calculation of the space coordinates is
made by,
\begin{eqnarray}
\int d {\vec k} \langle {\vec x}_1| {\vec k} \rangle \langle {\vec k}|{\vec
x}_2 \rangle = \delta({\vec x}_1-{\vec x}_2),
\end{eqnarray}
and the only one point $\delta {\vec x}=0$ satisfies the condition and 
the probability get a contribution from only the point $\delta {\vec
x}=0$. The rotational invariant  three dimensional space is compact
but the Lorentz invariant  four dimensional space is non-compact.   
This difference is important for   the reason why the relativistic
system has a peculiar  property of the interference.

It is worthwhile to clarify the difference of the space-time dependent
probability of the present work  with the normal scattering
probability defined at $t=\pm \infty$ here. 
The normal scattering
amplitude is defined from the overlap between the in-state at
$t=-\infty$ and out-state at $t=\infty$, and the space and time
coordinates are 
integrated from $-\infty$ to $\infty$ and the energy and momentum of the
final state is the same as that of the initial state.
Hence the momentum  of the muon or the pion in the final state of the ordinary 
scattering  experiments are bounded due to the energy momentum conservation.  
So the infinite momentum is not included in the muon or pion of the 
final state.  However the amplitude and probability at the finite time
and their behaviors at the finite time 
are  not computable  in the ordinary S-matrix. 

In our method it is possible to compute the amplitude and probability at
the finite time and  space. The  energy
and momentum conservation does not hold for these quantities and the infinite
momentum state of the muon and pion are included. These states of the infinite
momentum  give the finite
contribution 
to the time dependent probability but do not contribute to the 
cross section measured at infinite
distance.   The important informations are obtained from the wave packet
formalism that  are not calculable in the standard scattering amplitude.
Hence our calculation does not contradict with the ordinary calculation of the
S-matrix in momentum representation but has the advantage of giving new 
informations.

In our calculation, Lorentz invariance is  one important ingredient. 

The characteristic small phase of the relativistic wave packet shows 
macroscopic interference of the neutrino. Although this
result should be applied in high energy
region, it would be interesting to see if this effect is
found in ground experiments and others. Depending on the mass value,
the phenomenon we have discussed in this paper may be relevant to 
short base line experiments, long base line experiments, and atmospheric
neutrino experiments and others.

The oscillation phenomenon of the present work is sensitive to small
mass, hence the same mechanism would work if there exists a very light
particle. A possible candidate of light particle is  axion. Axion might
show a peculiar oscillation if it exist.

In this paper we ignored the effects of the pion life time and the pion 
mean free path in studying the higher order quantum effects. We will
study these problems  and other large scale physical phenomena 
of low energy neutrinos in subsequent papers.

    
\section*{Acknowledgements}
One of the authors (K.I) thanks Dr. Nishikawa for useful discussions on 
the near detector of T2K experiment, Dr. Asai, Dr. Mori, and Dr. Yamada
for useful discussions on interferences. This work was partially supported
by a Grant-in-Aid for Scientific Research(Grant No. 19540253 ) provided
by the Ministry of Education,
Science, Sports and Culture,and a Grant-in-Aid for Scientific Research on 
Priority Area ( Progress in Elementary Particle Physics of the 21st
Century through Discoveries  of Higgs Boson and Supersymmetry, Grant 
No. 16081201) provided by 
the Ministry of Education, Science, Sports and Culture, Japan. 
\\
{}

\appendix
\def\thesection{Appendix \Alph{section}}
\def\thesubsection{\Alph{section}-\Roman{subsection}}

\section{Formula for neutrinos of three flavour}
There are three eigenstates of neutrinos and their mass difference squared are known
and mixing parameters are also known partly from flavour oscillations.
The wave packets have been studied in flavour oscillations in
\cite{Kayser,Giunti,Nussinov,Kiers,Stodolsky,Lipkin,Asahara}. We studied
neutrino spatial interference in this paper, which  is  unrelated
directly with these flavour oscillations. A unified treatment of
neutrino phenomena is possible and will be presented in a next paper. 

For the wave packets to overlap at the detector, two components of mass
eigenstates should arrive to the detector same time and should have
the same energy within wave packet uncertainties. Hereafter we assume
that these    conditions are satisfied and study the amplitude for 
 three neutrinos. General finite time amplitude  for  each flavour
 combination is
 expressed using  amplitudes of three mass eigenstates, $T(i,t_1,{\vec
 x}_1)$ as  
\begin{eqnarray}
T_{\alpha,\beta}=\int_{T_0}^T dt_1 d{\vec x}_1\sum_i U_{\alpha
 i}T(i,t_1,{\vec x}_1)U^{\dagger}_{i \beta},
\end{eqnarray} 
where $U_{i\alpha}$ is a unitary matrix which connects flavour eigenstate
$\alpha$ to the mass eigenstate $i$. Time dependent neutrino production  
amplitude $T(i,t_1,{\vec x}_1)$ was  given before 
and is substituted into the finite time probability. We have, after the coordinates and other
variables are integrated,  
\begin{eqnarray}
& &|T_{\alpha,\beta}|^2=U_{\alpha i_1}(U_{\alpha i_2})^{*} \int_{T_0}^T dt_1
 dt_2T(i_1,t_1)
T(i_2,t_1)^{*} U^{\dagger}_{i_1 \beta}(U^{\dagger}_{i_2 \beta})^{*},\\
& &T(i_1,t_1)T(i_2,t_2)^{*}= N e^{i {m_{i_1}^2 \over E(p)} (t_1-T_0)-i {m_{i_2}^2
 \over E(p)} (t_2-T_0)}{1 \over t_1-t_2},
\end{eqnarray} 
where N is a constant.


\section{Operator product expansion and  new universal term}\label{App:OPE}
\subsection{Pion decay}
Neutrino detection amplitude and probability we have discussed is
understood from operator product expansions at the light cone~region 
$(x_1-x_2)^2 \approx 0$\cite{Wilson-OPE}. 
 
In Eq.($\ref{probability}$), the total probability is given by the
integral of the space time coordinates of the weak Hamiltonian and is
invariant under the translation of space and time. Due to the
translational invariance, the energy and momentum of the final state are
the same as those  of the initial state.  Now we
interchange the order of the summation of the coordinates and final
states and obtain the time dependent probability.
The time dependent probability which is obtained by summing the final 
state first is given in Eq.($\ref{probability2}$). This probability has
two components, invariant term  and non-invariant term under the
translation in time. The former is
the T-linear term and the latter is the T-oscillating term.

The translational invariant term gets the contribution from
 the final states that has  the same energy as the that of the initial
state. But  the energy of the final states that contributes to the
 latter is not necessary be the same as that of the initial state. 
 Hence the states of the infinite momentum could appear in
the intermediate state for the T-oscillating term and give the
 finite contribution to the probability at the finite time although they 
do not contribute to the probability at the infinite time. The states of 
infinite momentum contributes only to T-oscillating term and we
 estimate its magnitude based on the operator product expansion at the
 light cone region.    

The
functions $\Delta_\text{muon}(\delta t,\delta {\vec x})$ and
$\Delta_\text{pion}(\delta t,\delta {\vec x})$ at the light cone region, 
\begin{eqnarray}
\lambda =(\delta t)^2-(\delta {\vec x})^2=0
\end{eqnarray}
are the  expectation values of the products of the muon field and the
expectation value of  the pion field 
\begin{eqnarray}
& &\langle 0|\mu(x_1) \bar \mu(x_2) | 0 \rangle,\\ 
& &\langle \alpha_\text{in},\beta_\text{out} |\phi_\pi(x_1) \phi_\pi(x_2)
 |\alpha_\text{in},\beta_\text{out}  \rangle,
\end{eqnarray} 
where $\alpha_\text{in}$ includes the proton  and the target state and
$\beta_\text{out}$ includes many pions and other particles in the final
state.

The muon correlation function was studied in
Eq.~$(\ref{muon-correlation})$. 

\subsubsection{the normal  term}
Hereafter we study the pion correlation function. The pion
correlation function from Fig.~\ref{fig:pi-standard} gives the neutrino 
production probability that is determined by incoherent decay of produced 
pions and interference among the pions are negligible. The time 
dependence is given by 
$e^{i{m_{\pi}^2 \over E_{\pi}}(t_1-t_2) }$
and the angular velocity is too large to observe the oscillation. The
correlation function becomes 
\begin{eqnarray}
& &\Delta_\text{pion}(\delta t,\delta {\vec x})=\int d{\vec p} ~P({\vec p})e^{i{m_{\pi}^2 \over E_{\pi}}(t_1-t_2) }\\
& &=\langle |N_{\pi}|\rangle e^{i{m_{\pi}^2 \over < E_{\pi}>}(t_1-t_2) },
\end{eqnarray}
and in the ordinary pion energy of about $10~\text{GeV}/{c^2}$, the angular
velocity of this term corresponds to the oscillation length 
\begin{eqnarray}
L_0={ E_{\pi} \hbar c \over m_{\pi}^2},
\end{eqnarray}
which is too short for observation. The average of this term contributes
to the T-linear probability.   

\subsubsection{a new universal term}
\begin{figure}[t]
 \begin{minipage}{0.5\hsize}
  \begin{center}
  \includegraphics[scale=.3]{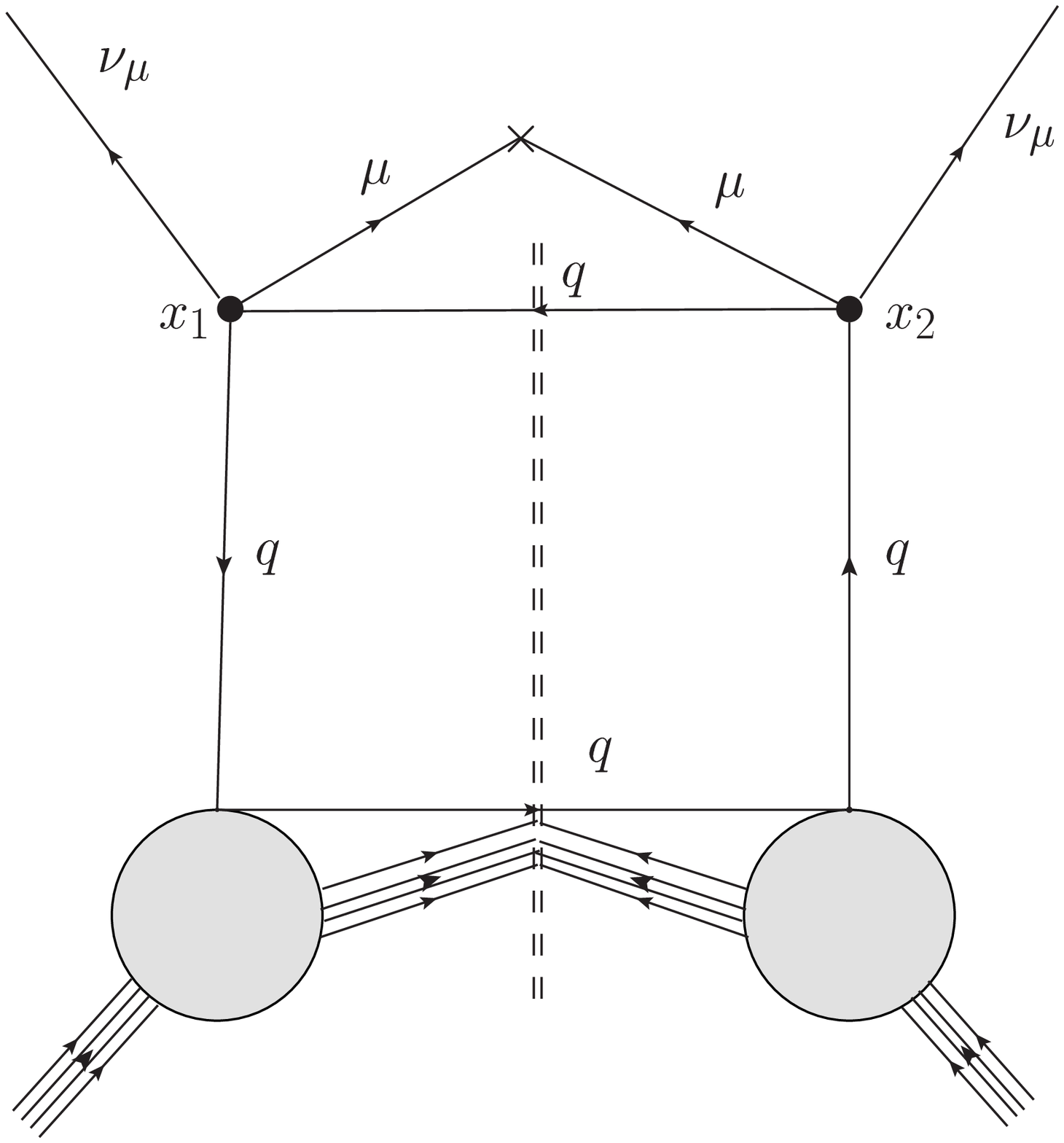}
  \end{center}
 \caption{The Feynman diagram of the anomalous term of the neutrino
  probability using quark fields is  given.}
 \label{fig:q-line}
 \end{minipage}
 \begin{minipage}{0.5\hsize}
  \centering
  \includegraphics[scale=.2]{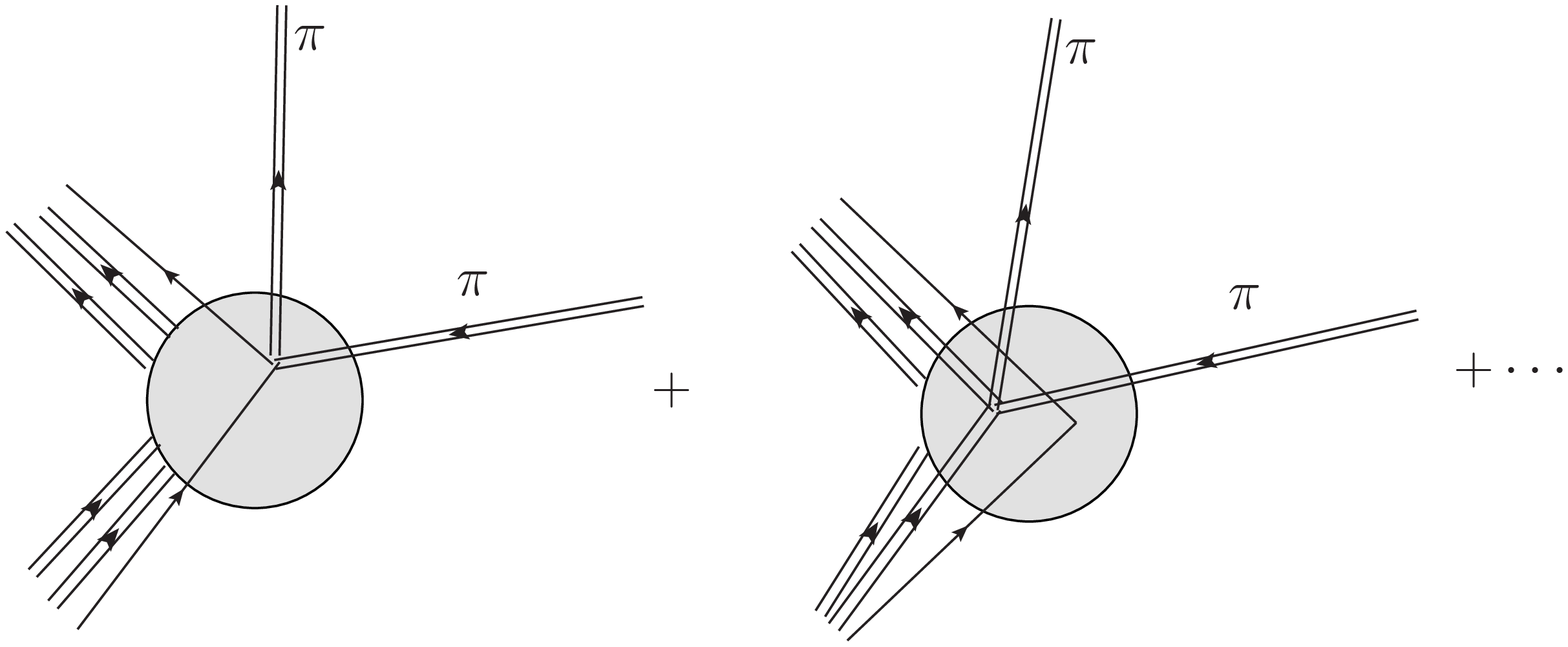}
  \caption{In the anomalous term of the neutrino probability, the hard pion
  couples   either with the pion or the proton in the hadronic parts.}
  \label{fig:pipi-scat} 
\end{minipage}
\end{figure}%
The energy of the final state  $\beta$ in the T-oscillating term is not
necessary the same as that of the initial state and infinite momentum
states can couple and gives  
finite contribution to the pion correlation function.  We estimate the 
effects of the infinite momentum hereafter.

We study the pion
correlation function that includes infinite momentum states in $\beta$.
They are  almost equivalent to Fig.~\ref{fig:pi-standard} except that
pions have infinite momenta and we write the
infinite momentum pion explicitly as 
Fig.~\ref{fig:pi-line}. We will see that this  gives a manifestly
invariant term
$F(\lambda)$ of Eq.$(\ref{pion-correlation})$. 

This correlation function is calculated with quark fields  as in 
the Feynman diagram of Fig.~\ref{fig:q-line} or with pion fields   
as in Fig.~\ref{fig:pi-line}.  In the former, QCD is applied. The
amplitudes in QCD has severe  infrared
divergence in time-like region and  quark propagator has no simple pole 
and has a cut at real energy axis as in QED discussed next.  We replace  the 
integration on the energy in Minkowski metric to the Euclidean four momentum
integration. This Wick rotation  is allowed if the amplitude is
analytic except cut along real axis. We assume that 
this holds and compute  the integral in  Euclidean metric. In Euclidean
metric there are no infrared divergence and quark dynamics  are effectively
described by meson dynamics. So we compute  the pion correlation
functions  using   pion propagators as in Eq.$(\ref{pion-correlation})$.  

Now we estimate the amplitude.
\begin{eqnarray}
& & F((x_1-x_2) )\\
& &=\int d^4p e^{-ip\cdot (x_1-x_2)} {1 \over p^2-m^2-i\epsilon}
{1 \over p^2-m^2+i\epsilon} \delta((p+p_f)^2-m^2)|T_\text{hadron}^1|^2 \nonumber\\
& &=\int d^4p ~e^{-ip\cdot (x_1-x_2)} {1 \over p^2-m^2-i\epsilon}{1 \over
p^2-m^2+i\epsilon}  {1 \over 2i\pi}\nonumber \\
& &\times ({ 1 \over (p+p_f)^2-m^2-i\epsilon}-{ 1 \over
 (p+p_f)^2-m^2+i\epsilon})|T_\text{hadron}^1|^2 \nonumber,
\end{eqnarray}
where $p_f=p_{\alpha}-p_{\beta}$. 
The momentum of the
pion which connect the coordinates $x_1$ and $x_2$ must be taken from
$-\infty $ to $\infty$. The integral is reduced to  
\begin{eqnarray}
& & F((x_1-x_2) )\\
& &=|T_\text{hadron}^1|^2{1 \over p_f (x_1-x_2)} 
{1 \over 64\pi^2}{1 \over m^2}+O\left({1
 \over \left\{p_f(x_1-x_2)\right\}^2}\right), \nonumber
\end{eqnarray}
for $p_f \neq 0$ in the Euclidean metric
space calculation. The infinite  momentum state of the pion at the 
coordinate $x_1$ enters the
hadronic part  $T_\text{hadron}^1$ and interacts with the one of pions or 
proton  and the pion of the infinite momentum goes out because 
other particles in the  hadronic part have finite momentum as
Fig.~$\ref{fig:pipi-scat}$. 

For $p_f=0$,  the
invariant term is computed as  
\begin{eqnarray}    
& &F\left({(x_1-x_2)}^2\right)\\
&=&g^2\int {d^4{p_{E}}\over  (2\pi)^4}e^{ip_E(x_1-x_2)}\left({1 \over
 p_E^2 +m^2}\right)^3 |T_\text{hadron}^1|^2.\nonumber
\end{eqnarray}
The integral in the right-hand side
\begin{eqnarray}
D((x_1-x_2)^2)= \int {d^4{p_{E}}\over  (2\pi)^4}e^{ip_E(x_1-x_2)}\left({1 \over
 p_E^2 +m^2}\right)^3
\end{eqnarray}
is invariant under the Lorentz transformation of $x_1-x_2$ and is a
function of $\lambda$. Hence we have 
\begin{eqnarray}
D(0)= \int {d^4{p_{E}}\over  (2\pi)^4}\left({1 \over
 p_E^2 +m^2}\right)^3={1 \over 64\pi^2}{1 \over m^2},
\end{eqnarray}
and  
\begin{eqnarray}    
& &F(0)\\
& &=g^2\int {d^4{p_{E}}\over  (2\pi)^4}
\left({1 \over
 p_E^2 +m^2}\right)^3 |T_\text{hadron}^1|^2  \nonumber \\
 & &={g^2 \over 64\pi^2} {1 \over m^2} |T_\text{hadron}^1|^2.             
\nonumber
\end{eqnarray}
Actually because two pion couples with the proton or pions and the total number
of the pions $N_{\pi}$ is larger than the number of the proton. Hence
the pion correlation function becomes 
\begin{eqnarray}    
F(0)
={g_{\pi N}^2+N_{\pi} g_{\pi\pi}^2 \over 64\pi^2} {1 \over m^2}
|T_\text{hadron}^1|^2,           
\end{eqnarray}
where $g_{\pi N }$ is the  pion Nucleon coupling strength and
$g_{\pi\pi} $ is the pion pion coupling strength,  

\subsection{Real pion decay}
When the real pion of the momentum $p_\text{pion}$ is  an initial state,
one loop correction to the pion correlation function is given by the
following integral,
 \begin{eqnarray}
& & \Delta_{\pi}({ p}_\text{pion},x_1-x_2)\\
& &=\int d^4q ~e^{-iq\cdot (x_1-x_2)} {1 \over p^2-m^2-i\epsilon}{1 \over
p^2-m^2+i\epsilon}\delta((p+q_\text{pion})^2) \nonumber\\
& &=\int d^4q ~e^{-iq\cdot (x_1-x_2)} {1 \over p^2-m^2-i\epsilon}{1 \over
p^2-m^2+i\epsilon}  {1 \over 2i\pi}\nonumber \\
& &\times ({ 1 \over (p+q_\text{pion})^2-m^2-i\epsilon}-{ 1 \over (p+q_\text{pion})^2-m^2+i\epsilon}).
\end{eqnarray} 
When the higher order correction is added and the infrared divergence is
avoided, the propagator becomes not to have a simple pole but a cut
along the real axis. Although it is a difficult problem to find out the
dynamics of infrared divergence, we simply make Wick rotation and
compute the Lorentz invariant term.
  
This integration is made in Euclidean metric space. We have then,
\begin{eqnarray}
& &\Delta_{\pi}({ p}_\text{pion},x_1-x_2)\\
& &={1 \over p_\text{pion}(x_1-x_2)} {1 \over 64\pi^2}{1 \over m^2}+O\left({1
 \over \left\{p_\text{pion}(x_1-x_2)\right\}^2}\right), \nonumber
\end{eqnarray}
which has a power term. 
\begin{figure}[t]
 \begin{minipage}{0.5\hsize}
  \centering
  \includegraphics[scale=.45]{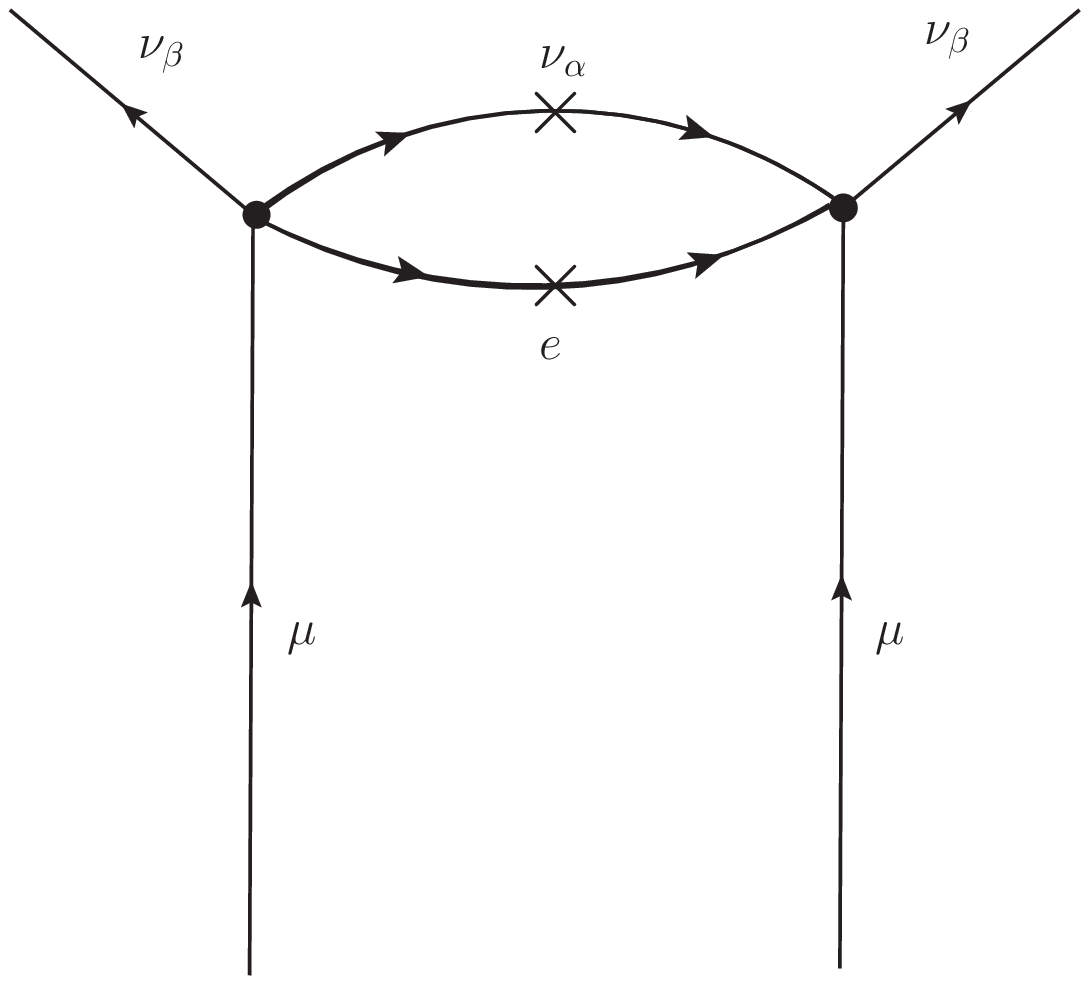}
  \caption{The diagram of the normal term in the neutrino probability 
in the muon decay.}
  \label{fig:mu-nogamma} 
 \end{minipage}\ \ \ 
   \begin{minipage}{0.5\hsize}
    \centering
    \includegraphics[scale=.45]{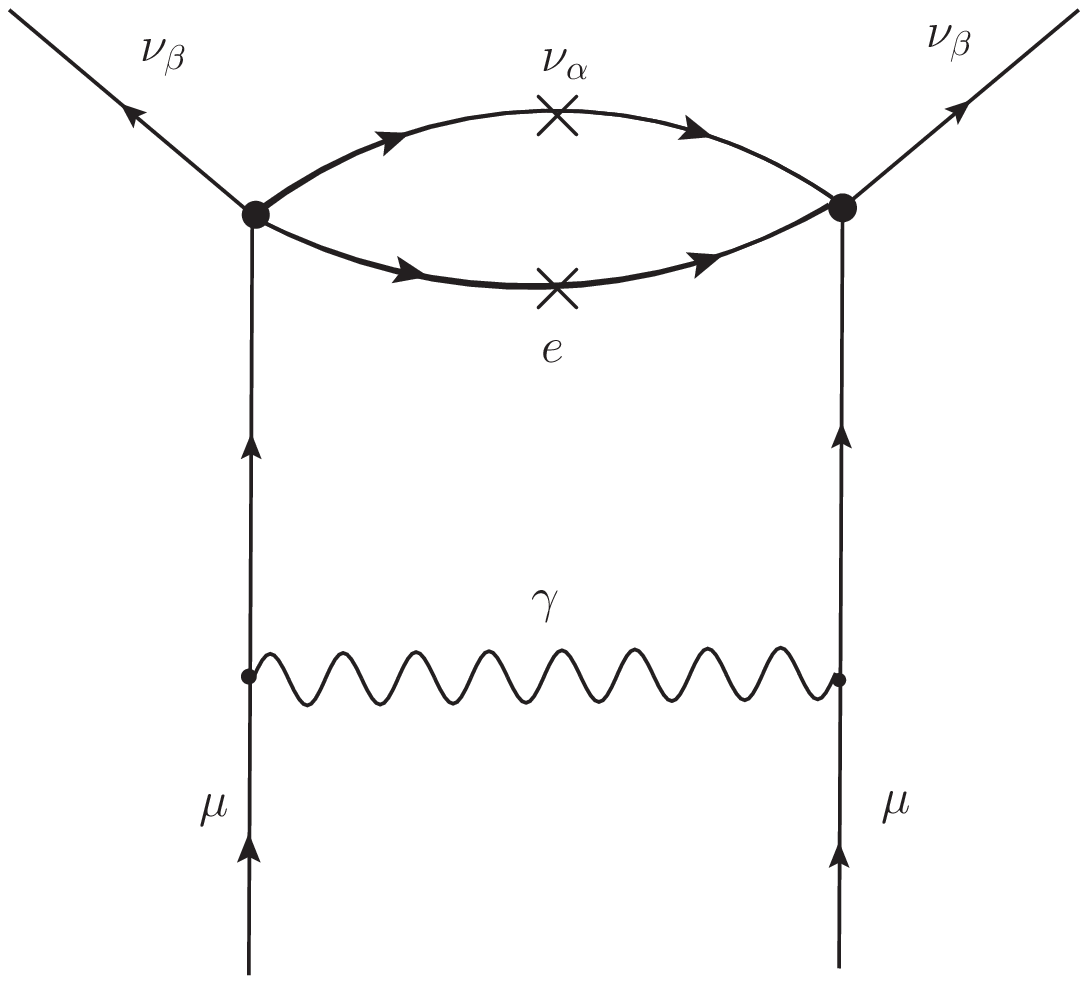}
 \caption{The diagram of the lowest order correction in the neutrino probability 
in the muon decay.} 
 \label{fig:mu-gamma}
   \end{minipage}
\end{figure}
\subsection{III muon decay}
 For the muon decays, higher order corrections  are  generated from QED  and 
the hard photon exchange term of Fig.~\ref{fig:mu-gamma} would give the invariant
term. This diagram has the infra-red divergence that is avoided  by
redefining one particle
charged state in the initial and final state in such way that is dressed
by soft photon. Charged field becomes not to have simple pole but
cut. So integration of the large momentum in Fig.~\ref{fig:mu-gamma} becomes equivalent
to that of the Euclidean metric. We estimate the Lorentz invariant
amplitude from the integration in the Euclidean metric integration and we
have then       
\begin{eqnarray}    
& &F_\text{muon}(x_1-x_2)\\
& &=e^2\int {d^4{p_{E}}\over  (2\pi)^4}e^{ip_E(x_1-x_2)}
({1 \over
 p_E^2 +m_{\mu}^2})^2{1 \over (p_E+q)^2} |T_\text{muon}^1|^2  \nonumber \\
 & &={e^2 \over 64\pi^2} {1 \over q(x_1-x_2)}{1 \over m_{\mu}^2}
  |T_\text{muon}^1|^2 +O({1 \over (q(x_1-x_2))^2}),            
\nonumber
\end{eqnarray}
where $q$ is the momentum of muon.
Thus higher order correction has a power correction term that is inversely 
proportional to the variable $q(x_1-x_2)$ but not a function of
$\lambda$. This term gives also
the oscillation of the frequency $m_{\nu}^2 \over E_{\nu}$. However
because of the factor ${1 \over q(x_1-x_2)}$ this oscillation appears in
the time derivative of the probability.
\end{document}